\documentclass[english,aps,prb,reprint,superscriptaddress,nofootinbib]{revtex4-2}
\usepackage[T1]{fontenc}
\usepackage[latin9]{inputenc}
\usepackage{verbatim}
\usepackage{amsmath}
\usepackage{amsthm}
\usepackage{amssymb}
\usepackage{graphicx}
\usepackage{esint}
\usepackage{textcomp}
\usepackage{caption}
\usepackage{color}
\usepackage{bm}
\usepackage{hyperref}
\makeatletter

\renewcommand{\fnum@figure}{FIG. \thefigure}

\makeatother

\usepackage{babel}

\begin{document}

\title{Stokes flow in an electronic fluid with odd viscosity}
	
\author{Yonatan Messica}
\affiliation{
	Department of Physics, Bar-Ilan University, Ramat Gan, 52900, Israel
}
\author{Alex Levchenko}
\affiliation{Department of Physics, University of Wisconsin-Madison, Madison, Wisconsin 53706, USA}

\author{Dmitri B. Gutman}
\affiliation{
	Department of Physics, Bar-Ilan University, Ramat Gan, 52900, Israel
}
\date{\today}

\begin{abstract}

We investigate the transition between elastic and viscous regimes  for time-reversal broken Weyl semimetals.
In these materials, Hall transport occurs through two parallel channels: the Fermi sea and the Fermi surface. The Fermi sea part remains unaffected by electron-electron scattering, whereas the Fermi surface is influenced by it. We model the disorder by dilute impenetrable spherical impurities. 
We analyze the flow  of  electronic fluid 
with a finite odd viscosity  in the presence of such disorder 
and compute the conductivity tensor.
We find that in the generic case of finite intrinsic conductivity, the Hall angle in the viscous regime is parametrically suppressed compared to the elastic regime. In the special case where the intrinsic conductivity vanishes, the ratio between the transverse and the longitudinal resistivities matches the ratio between the odd and even components of the viscosity tensor.
\end{abstract}

\maketitle

\textit{Introduction}. The interplay between solid-state physics and fluid dynamics is a two-way street. On the one hand, the hydrodynamic description provides deeper insights into the collective motion of electrons under external forces, capturing their behavior in the viscous regime. This behavior qualitatively differs from that of noninteracting electrons, exhibiting remarkable nonlocal features \cite{Bandurin2016}, and higher than ballistic conduction through microconstrictions \cite{Kumar2017}. 
On the other hand, solid-state systems introduce novel hydrodynamic problems that are inconceivable for conventional fluids. These systems are governed by hydrodynamic equations fundamentally distinct from those describing traditional fluids. A striking example is the presence of anomalous terms, which can be traced back to the Berry curvature in the conduction band of electrons \cite{Xiao2010}.

In this Letter, we study the transition between elastic and hydrodynamic transport regimes in time-reversal broken Weyl semimetals.
An unusual aspect appears in this problem due to an odd viscosity term~\cite{Avron1995,Avron1998,Fruchart2023}. Such terms are permitted when the time-reversal symmetry is broken. They arise, for example,  in active matter systems of self-spinning rotors~\cite{Banerjee2017,Markovich2021,Hosaka2023,Lier2023,Everts2024,Hosaka2024}.
In the condensed matter setting odd viscosity was studied in the gapped state of the quantum Hall effect ~\cite{Read2011,Haldane2009,Hoyos2012}. 
For noninteracting electrons  this quantity is robust, provided the rotational invariance is preserved, and  can be probed by measuring the finite-$q$ part  of the Hall conductivity \cite{Hoyos2012,Bradlyn2012}.
Hall viscosity was also analyzed in viscous electronic hydrodynamics in the presence of an external magnetic field~\cite{Alekseev2016,Scaffidi2017,Delacretaz2017,Holder2019,Afanasiev2022}.

In two dimensions, the odd viscosity does not modify the flow of an incompressible fluid under the no-slip boundary conditions \cite{Ganeshan2017}. As a result, the odd viscosity does not affect the resistivity in the thermodynamic limit for systems with diffusive scatterers \cite{Alekseev2023,Gornyi2023}. 
In contrast, as we demonstrate in this work, for  three-dimensional systems, 
odd viscosity has a profound effect on the transport in the viscous hydrodynamic regime. 

We focus on the  3D electronic fluid in the gapless state formed in Weyl semimetals with broken time-reversal symmetry. Unlike standard magnetohydrodynamics, there is no average magnetic field or a corresponding Lorentz force. This absence makes the effects associated with anomalous transport in semimetals more prominent.
We consider a model of impenetrable spherical impurities of radius $R$ randomly distributed with density $n_{\text{imp}}$.
This model of impurities has been used  to study transport in  Weyl semimetals \cite{Messica2023}, graphene \cite{Krebs2023}, 
and to model a  polaron-like state, commonly referred to as an electron bubble in $^3$He-A~\cite{Ikegami2013}. Additionally, this model can approximate a screened potential due to charged impurities. For low carrier concentrations, the range of such potential can be significantly larger than the atomic length scales \cite{Kwon2006}.

We briefly mention that electron hydrodynamics in Weyl semimetals have also attracted attention due to the chiral nature of the Weyl electrons. Applying an electric field parallel to an external magnetic field pumps charge from one Weyl node to another, a phenomenon known as the chiral anomaly \cite{Son2013}. Assuming slow intervalley relaxation, this leads to multi-mode hydrodynamics describing two interacting fluids, or, a density mode and an imbalance mode \cite{Landsteiner2015, Lucas2016, Gorbar2018, Sukhachov2018}. In the present work, we consider no external magnetic field and thus stay within single-mode hydrodynamics, focusing on the unique effects of the odd viscosity.

We consider the case where the chemical potential is far from the Dirac points, and
explore the transition between the elastic and viscous regimes as the temperature increases. We start with the viscous hydrodynamic limit
where the electron-electron scattering length is much smaller than the electron-disorder scattering length, $l_{\rm{ee}}\ll l_{\rm imp}$. In this regime, the electrons are in a local thermal equilibrium and can be described by hydrodynamic equations \cite{Gorbar2018,Narozhny2019,Narozhny2022,Fritz2024}.

In Cartesian coordinates, the Navier-Stokes equation reads \cite{Landau2003}
\begin{equation}
\label{Navier-Stokes}
\frac{\partial v_i}{\partial t}+ v_j\frac{\partial v_i}{\partial x_j} =-\frac{1}{mn}\frac{\partial p}{\partial x_i}+\frac{1}{mn}\frac{\partial \sigma_{ij}}{\partial x_j}.   
\end{equation}
Here ${\bf v}$ is the velocity field, $n$ is the particle density of the electronic fluid, $m$ is the effective mass, and $p$ is the pressure in the fluid. The viscous part of the stress tensor is given by 
\begin{equation}
\sigma_{ij}(v)=\frac{1}{2}\eta_{ijkl}\left(\frac{\partial v_k}{\partial x_l} +\frac{\partial v_l}{\partial x_k} \right).
\end{equation}
The stress tensor is symmetric, i.e.,  $\sigma_{ij}=\sigma_{ji}$. The viscosity tensor $\eta_{ijkl}$ can, in turn,  be divided into two parts: 
\begin{equation}
\eta_{ijkl}=\eta_{ijkl}^e+\eta_{ijkl}^o\,.
\end{equation}
The even part $\eta_{ijkl}^e$ is a fully symmetric tensor, and it is the part that is considered in standard hydrodynamics~\cite{Landau2003}. 
However, in systems  with broken time-reversal symmetry, the viscosity tensor may also 
have an antisymmetric part~\cite{Avron1995,Avron1998}
\begin{equation}
\eta_{ijkl}^o=\eta_{jikl}^o=\eta_{ijlk}^o=-\eta_{klij}^o,
\end{equation}
often referred to as odd viscosity.
These components capture distinct physical effects in the fluid's stress response to deformation. The even part gives rise to parallel friction between layers of the fluid moving relative to each other (shear flow). The odd part of the viscosity exerts asymmetric perpendicular forces (i.e., normal to the layer's boundary) in response to a shear flow profile. The resulting stress tensor can be decomposed into even viscosity and odd viscosity parts,
 $\sigma_{ij}(v)={\sigma}_{ij}^e(v)+ {\sigma}_{ij}^o(v)$.
The influence of the odd viscosity is particularly interesting in the context of anomalous Hall transport. Indeed, in the absence of an external magnetic field, which typically dominates Hall transport by exerting the Lorentz force on the electrons, other mechanisms contributing to Hall transport become more prominent~ \cite{Haldane2004,Nagaosa2010}.

\textit{Model and odd Stokes flow}. We now focus on the electronic fluid in a Weyl semimetal (WSM) with broken time-reversal symmetry \cite{Yan2017,Armitage2018,Gooth2018,Jaoui2018} \footnote{We note that since we are interested in the hydrodynamic odd viscosity, time-reversal symmetry needs to be broken for the low-energy model describing the vicinity of the Weyl nodes. A linear dispersion model for a Weyl node has an emergent time-reversal symmetry, defining the inversion of momentum with respect to the node's position in momentum space. Thus, the low-energy Hamiltonian for the system considered here needs to break this emergent symmetry, as can be done with a finite tilt of the Weyl node or by including non-linear corrections to the dispersion relation.}. For simplicity, we consider a minimal model of two Weyl nodes with opposite chirality \cite{Burkov2011}.
The breaking of the time-reversal symmetry defines a spatial direction that we denote as
$\hat{\mathbf{z}}$. For simplicity, we assume that the
single-particle spectrum of the electrons is symmetric about rotation around this axis. 
Based on symmetry arguments one can construct  the odd component of the viscosity tensor as 
\begin{equation}
\label{oddviscosity}
\eta_{ijkl}^o=\frac{\eta^o}{2}
\left(\delta_{ik}\epsilon_{jl}+\delta_{jl}\epsilon_{ik}+\delta_{jk}\epsilon_{il}+\delta_{il}\epsilon_{jk}\right),
\end{equation} 
where we define the reduced antisymmetric Levi-Civita symbol $\epsilon_{jl}\equiv\epsilon_{jl3}$.
We consider an incompressible fluid
\begin{equation}
\nabla\cdot  {\bf v}=0\,.
\end{equation}
For this model, the  divergence of the viscous part of the stress tensor  has  the following form
\begin{equation}
\label{NS}
\frac{\partial\sigma_{ij}}{\partial x_j}=
\eta^e\Delta v_i+\frac{1}{2}\eta^o\left(\partial_i \partial_j\epsilon_{jk}v_k+\Delta\epsilon_{ik}v_k\right).
\end{equation}

To account for interaction with impurities, we now turn to the problem of the sphere of radius $R$ moving with respect to the fluid with relative velocity $-\bf{u}$. It is convenient to pass into the reference frame of the sphere, such that the fluid at infinity has a constant velocity ${\bf v}(r\rightarrow \infty)\simeq {\bf u} \equiv u\hat{\mathbf{x}}$.
This reference frame is often more natural in the condensed matter setting, where the sphere mimics a fixed, finite-sized impurity. In this reference frame, the velocity field is static in the steady state. Assuming that the characteristic Reynolds number is small, i.e. ${\rm Re} \equiv u mn R /\eta^e\ll 1 $, the Navier-Stokes equation for the velocity field can be approximated as
\begin{equation}
\label{NS_linearized}
\frac{\partial p}{\partial x_i}-
\eta^e \Delta v_i-\frac{1}{2}\eta^o\left(\partial_i \partial_j\epsilon_{jk}v_k+\Delta \epsilon_{ik}v_k\right)=0.
\end{equation}
Applying  the  curl to  Eq. (\ref{NS_linearized}) nullifies the gradient terms, and we arrive at   
\begin{equation}
\label{NSlinearized_curl}
\eta^e\Delta \nabla \times {\bf v}+\frac{1}{2}\eta^o \Delta \partial_z {\bf v}=0\,.
\end{equation}
This equation is supplemented with the standard no-slip boundary condition, which requires the velocity to vanish on the surface of the sphere: ${\bf v}(r=R,\theta,\phi)=0$. Note that the second term in Eq.~\eqref{NS} is proportional to the odd viscosity and is absent in time-reversal (TR) symmetric materials.

We assume that $\gamma\equiv \eta^o/\eta^e\ll 1$ and proceed to compute the velocity field by a perturbative expansion   
around the Stokes solution
\begin{equation}
\label{perturbative}
 {\bf v}(r,\theta,\phi)={\bf v}_{0}+\gamma {\bf v}_1+\gamma^2{\bf v}_2+\dots.
 \end{equation}
In spherical coordinates, with the $z$-axis aligned along the direction of the time-reversal symmetry breaking, the Stokes solution reads \cite{Landau2003}:
\begin{align}\label{vStokes}
 &{\bf v}_0(r,\theta,\phi)-{\bf u}=\frac{u}{4}\bigg[2\hat{\mathbf{r}}\sin \theta \cos \phi \left(\frac{R^3}{r^3}-\frac{3R}{r}\right)-\nonumber
 \\
 &-\hat{\bm{\theta}}\cos\theta\cos\phi\left(\frac{R^3}{r^3}+\frac{3R}{r}\right)+ 
 \hat{\bm{\phi}}\sin\phi\left(\frac{R^3}{r^3}+\frac{3R}{r}\right)\bigg].
 \end{align}
 
\begin{figure*}[t!]
    \includegraphics[width=0.325\linewidth]{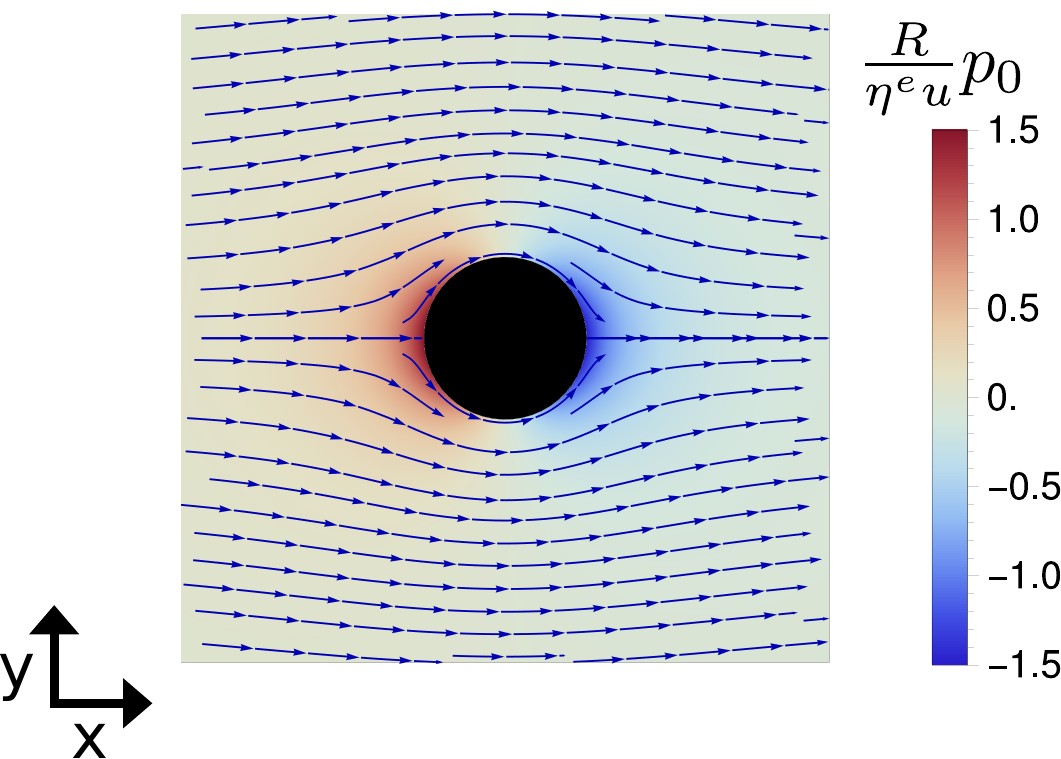}
     \includegraphics[width=0.325\linewidth]{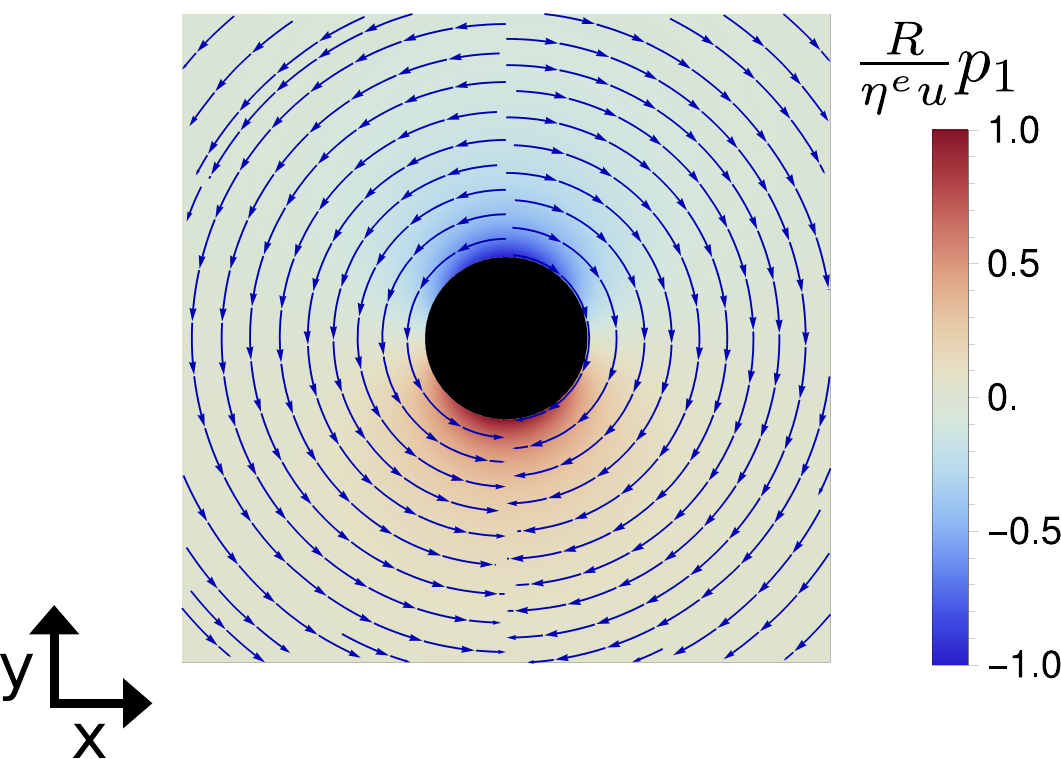}
     \includegraphics[width=0.325\linewidth]{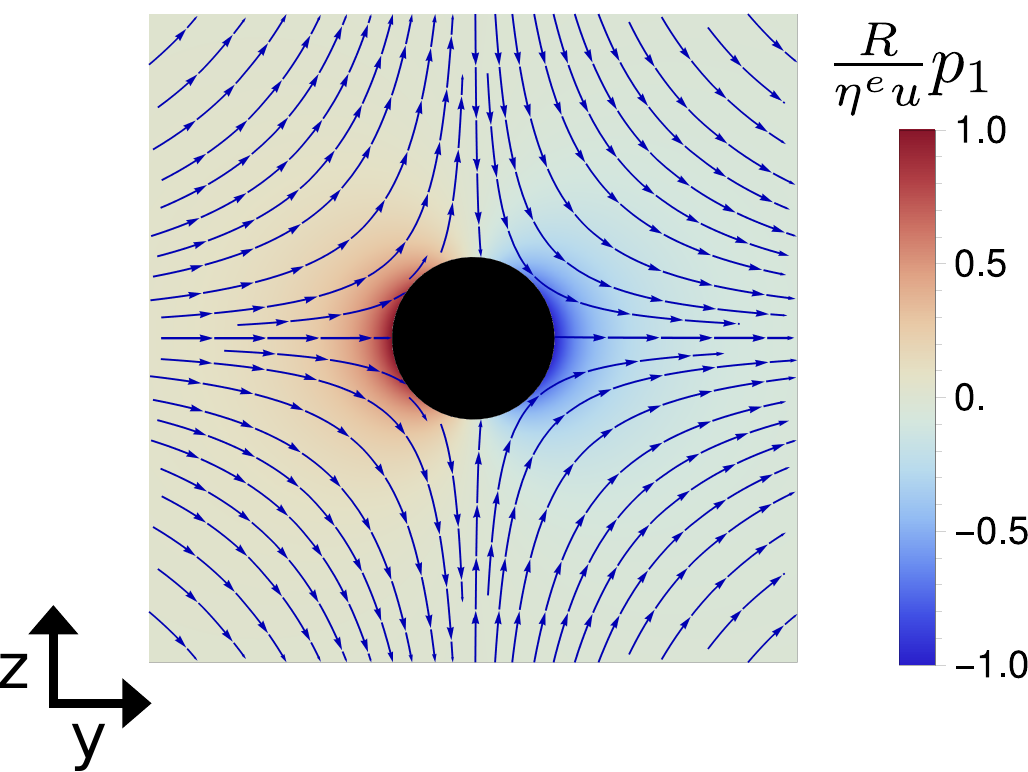}
    \caption{(Left): Streamlines of the velocity $v_0$ superimposed  with the pressure $p_0$ in the x-y plane. (Center): Streamlines of the velocity $v_1$ superimposed with the  pressure $p_1$ in the x-y plane. (Right): Streamlines of the velocity $v_1$  superimposed  with the  pressure $p_1$ in the y-z plane.}
    \label{fig:streamlines}
\end{figure*}

As the next step, we iterate this solution once in Eq.~(\ref{NSlinearized_curl}) and compute the first correction to the velocity field. 
Following the steps outlined in the Supplemental Material, we find the correction to the velocity field in the form 
 \begin{equation}\label{v1}
{\bf v}_1=\frac{3u}{16}\left(\frac{R}{r}-\frac{R^3}{r^3}\right)
\left(\hat{\bm{\theta}}\cos\theta\sin\phi+\hat{\bm{\phi}}\cos\phi\cos2\theta\right).
 \end{equation}
 This velocity field reproduces the results of Refs. \cite{Hosaka2023, Everts2024, Lier2024} in the limit of small $\gamma$ and Reynolds number. It is instructive to compare the correction to the velocity field with Eq.~(\ref{vStokes}). The flow described by ${\bf v}_1$ is purely tangential and contains quadrupolar terms in its angular dependence.
The streamlines of the velocity fields ${\bf{v}}_0$ and ${\bf{v}}_1$ 
are depicted in Fig. \ref{fig:streamlines}. 

We perform an analogous expansion for the pressure field
\begin{equation}
p=p_0+\gamma p_1+\dots\,, 
\end{equation}
and substitute this into Eq. (\ref{NS_linearized}).
After solving this equation one finds
\begin{eqnarray}&&
p_0({\bf r})=p_\infty-\frac{3}{2}\eta^e u\frac{R}{r^2}\sin \theta \cos \phi\,, \nonumber \\&&
p_1({\bf r})=-\frac{9}{8}\eta^e u\frac{R}{r^2}\sin\theta\sin\phi\,.
\end{eqnarray}
Here  $p_0$ reproduces the  known result for a fluid with  even viscosity, 
and $p_1$ is the pressure induced by an odd viscosity;   $p_\infty$ is the pressure far away from the sphere. The $p_0({\bf r})$ part is a dipole oriented along $\hat{x}$ axis and $p_1({\bf r})$ is a dipole oriented along $\hat{y}$ axis. The pressure profiles are visualized in Fig. \ref{fig:streamlines}. It is worth comparing the positions of the maximal and minimal values for the two pressure terms, shown in Table \ref{table:max_min_locations}.

\begin{table}[h!]
\centering
\begin{tabular}{|c|c|c|}
\hline
\textbf{Pressure} & \textbf{Maximum location} & \textbf{Minimum location} \\ \hline
$p_0$ & $-x$-axis & $x$-axis \\ \hline
$p_1$ & $-y$-axis & $y$-axis \\ \hline
\end{tabular}
\caption{Extremal points for $p_0$ and $p_1$.}
\label{table:max_min_locations}
\end{table}

The pressure induces a force that acts on the sphere (see the Supplemental Material for the detailed calculation)
\begin{equation}
\label{Fpressure}
{\bf F}^{p}=-R^2\int d\Omega\, \hat{\mathbf{r}}(\theta,\phi)p(R,\theta,\phi)=\pi \eta^e u R \left( 2  \hat{\mathbf{x}} + \frac{3}{2}\gamma \hat{\mathbf{y}} \right),
\end{equation}
where $\hat{\mathbf{r}}=\sin\theta\cos\phi \hat{\mathbf{x}}+\sin \theta\sin\phi \hat{\mathbf{y}}+\cos \theta \hat{\mathbf{z}}$ is the unit vector normal to the surface of the sphere and the integration is over the solid angle $d\Omega=\sin\theta d\theta d\phi$.
Another contribution to the force comes from the viscous part of the stress tensor. 
The force acting on a unit area of the sphere in the direction $\hat{\mathbf{e}}_i$ is given by
$F_i^\sigma=\sigma_{i,k}\hat{r}_k\,$. In spherical coordinates, it reads
$F_i^\sigma=\sigma_{r,r}\hat{r}_i+\sigma_{\theta,r}\hat{\theta}_i+\sigma_{\phi,r}\hat{\phi}_i$.
The total force coming from the viscous tensor is therefore
\begin{subequations}
\begin{equation}
{\bf F}^\sigma=\int d\Omega
\sum_{i=1}^3\hat{\mathbf{e}}_i F^\sigma_i= \int d\Omega
\sum_{i,k=1}^3\hat{\mathbf{e}}_i \hat{r}_k \sigma_{ik}(\mathbf{v}),
\end{equation}
\begin{equation}
\label{stress_tensor_decomposition}
\sigma_{ij}({\bf v}) = \sigma_{ij}^e({\bf v}_0)+\sigma_{ij}^e({\gamma\bf v}_1)+\sigma_{ij}^o({\bf v}_0).
\end{equation}
\end{subequations}
The first term in Eq. (\ref{stress_tensor_decomposition}) together with the pressure term $p_0({\bf r})$ in Eq. (14) give rise to the standard Stokes force~\cite{Stokes1851}
\begin{equation}
\label{standard_Stokes}
{\bf F}_{\parallel}=6\pi\eta^e Ru\hat{\mathbf{x}}.
\end{equation} 
This force acts in the direction opposite to the velocity of the sphere. 

Odd viscosity does not influence the force acting on a sphere in the ${\hat x}$-direction at the level of linear response. Indeed, due to the dissipationless nature of the odd viscosity, it does not generate any Joule heat, which equals the work done by the force, i.e., ${\bf F}^\sigma\cdot {\bf u}=0$. Our calculations are consistent with this condition. Another sanity check involves computing the torque exerted on the sphere by viscous forces. The lack of dissipation again dictates the absence of such a torque from odd viscosity terms. A finite torque would cause the sphere to spin, leading to energy transfer between the fluid and the sphere via odd viscosity terms. It is worth mentioning that such processes are possible when the problem is studied beyond the linear response regime \cite{Lier2024}.

We now combine the results for the pressure gradients 
and the viscous stress associated with the odd part of the viscosity.
Note that these forces act in opposite directions. This resembles the situation of the negative voltage drop due to the vortex formation in the viscous flow \cite{Levitov2016}. In our problem, the competition exists only in the transversal direction, while in the longitudinal direction, both forces are aligned. 
Combining the pressure and viscous stress forces, one finds
the total odd (Hall) component of the force acting on the sphere 
\begin{equation}
\label{force_total}
{\bf F}_\text{Hall}=-\frac{3\pi}{2}\eta^oRu\hat{\mathbf{y}}\,.
\end{equation}
This result agrees with the recent findings on microswimmers in an odd fluid \cite{Hosaka2023, Lier2024}. 
In the condensed matter setting, the moving sphere can represent a strong finite-size impurity.  In the case where the density of impurities is low, they interact with fluid independently, and the force described by Eq. (\ref{force_total}) is precisely the force that the fluid exerts on an individual impurity.

\textit{Electric transport in an odd viscous fluid}. We proceed to discuss the implications of the transition from elastic scattering to viscous flow for electric transport in a WSM.
The transport is carried in two parallel channels: 
the Fermi surface states (extrinsic channel) and the filled states,
\begin{equation}
\label{intrinsic_extrinsic}
{\bf j}={\bm \sigma}^{\rm ext}{\bf E}+{\bm \sigma}^{\rm int}{\bf E}.
\end{equation}
The latter is often referred to as the intrinsic contribution. 
It is controlled by the Berry curvature of the occupied electronic Bloch states,
and for a simple model, it is given by the sum of Weyl node dipoles
${\sigma}^{\rm int}_{xy} = e^2/ \left( 2\pi h \right) \sum_j \Delta^j_z$
\cite{Burkov2011, Yang2011, Burkov2014}, where  ${\bf{\Delta}}^j$  is the
distance in momentum space between the Weyl nodes belonging to the $j$-th dipole.
Because the involved states are located in the Fermi sea, they do not participate in real transitions and are thus unaffected by electron scattering. Therefore, the intrinsic contribution to Hall conductivity is robust and remains 
the same in the elastic and viscous regimes. The Fermi-surface contribution, on the other hand, changes.

At low temperatures, the e-e collision length $l_{\text{ee}}$ exceeds  the momentum relaxation  length due to the disorder scattering 
($l_{\rm ee} \gg l_{\rm imp}$).
For this model, the momentum relaxation rate due to static disorder can be estimated as $\tau^{-1}_{\text{imp}}\simeq v_{\text{F}}n_{\text{imp}}R^2$, where $v_F$ is the Fermi velocity (in this segment, we denote by "$\simeq$" equality up to a multiplicative factor of order unity).
In this regime, momentum relaxation is achieved by uncorrelated electron scattering, and in the leading order the extrinsic part of 
the conductivity is given by the Drude formula
\begin{equation}
  \sigma^{\text{Drude}}_{xx}\simeq 
  \frac{e^2}{p_FR^2} \frac{n}{n_{\rm imp}},
\,\,\sigma^{\text{Drude}}_{xy}\simeq \frac{\sigma^{\text{Drude}}_{xx}}{\epsilon_F\tau_{\rm imp}}\,,
\end{equation}
where $p_F$ is a Fermi momentum, and $\epsilon_F$ is the Fermi energy.
The corresponding Hall angle in this regime equals
\begin{equation}
\tan \theta_H=\frac{\sigma_{xy}}{\sigma_{xx}} \simeq
\frac{\sigma_{xy}^{\rm int}}{\sigma^{\rm Drude}_{xx}}
+\frac{1}{\epsilon_F\tau_{\rm imp}}\,.
\end{equation}
As the temperature increases, the scattering length $l_{\rm ee}$ decreases. When it becomes comparable to the size of the impurities ($l_{\rm {ee}} \sim R$), the scattering of the electron of the impurity can no longer be separated from the e-e scatterings and the system reaches the viscous regime~\cite{Hruska2002,Guo2016}. In this limit, there is no momentum relaxation due to the scattering of individual electrons by the impurity. Instead, it is an interaction between an impurity and the surrounding electrons. 
For $l_{\rm ee} \ll R$ the momentum relaxation is accounted for by the friction between the electronic fluid and the spheres, analyzed above. Assuming that the impurities are dilute $(n_{\rm imp}^{-1/3} \gg R)$,
we can rely on Eqs. (\ref{standard_Stokes}) and (\ref{force_total}) to compute the resistance. We note that for dilute impurities, the length scales satisfy $R\ll l_{\rm{imp}}$. Thus, the condition for momentum relaxation due to viscosity, $l_{\rm {ee}} \ll R$, is stricter than the condition for the onset of hydrodynamic flow, $l_{\rm {ee}} \ll l_{\rm{imp}}$ (see the Supplemental Material for a detailed discussion, which includes Refs. \cite{Sinitsyn2006,Atencia2022,Muller2009,Yudson2007,gantmakher1987carrier,Principi2015,Kiselev2019,Moessner2019}).

For steady-state flow, the force density exerted on a fluid element by an external electric field $en\mathbf{E}$ should be balanced with the friction from the Stokes force $n_\textrm{imp} \mathbf{F}$, with $\mathbf{F}=\mathbf{F}_{\parallel} + \mathbf{F} _{\textrm{Hall}}$ being the force from a single impurity. This fixes the value of the hydrodynamic velocity $u\simeq enE/n_{\text{imp}}\eta^eR$, and determines the charge current density, $\mathbf{j}=en\mathbf{u}$. Employing the kinetic expression for the viscosity $\eta^e\simeq n p_{\text{F}}l_{\text{ee}}$ \cite{LL-V10},
one can read off the corresponding conductivity tensor $\mathbf{j}=\bm{\sigma}\mathbf{E}$. Its longitudinal component is given by 
\begin{equation}
 \sigma^{\text{Stokes}}_{xx} \simeq \sigma^{\rm Drude}_{xx}
 \frac{R}{l_{\text{ee}}}.
\end{equation}
Because $R/l_{\text{ee}} \gg 1$ in the viscous regime, the longitudinal conductivity is parametrically larger than in the elastic regime. Moreover, in the Fermi liquid regime $l_{\text{ee}}\propto1/T^2$, implying that the viscous resistivity is of the insulating sign, $\partial_T\sigma_{xx}^{\rm Stokes}>0$, which is a manifestation of the Gurzhi effect \cite{Gurzhi1963}. 

The Hall component of the force due to the odd viscosity leads to the appearance of an electric field in the transverse direction relative to the flow. From the force balance condition, we get $E_y \simeq (\eta^o/\eta^e)E_x$. 
Employing Eq. (\ref{intrinsic_extrinsic}) and demanding that $j_y^{\rm ext}=0$ we get the relation
$E_y=E_x({\sigma}^{\rm ext}_{xy}/{\sigma}_{xx}^{\rm ext})$.
This general argument implies that in the viscous regime, the extrinsic parts of the conductivity  and viscosity tensors are proportional, i.e., ${\sigma}^{\rm ext}_{xy}/{\sigma}_{xx}^{\rm ext}
\simeq \eta^o/\eta^e$.
The corresponding Hall angle is therefore
\begin{equation}
\label{theta_H viscous}
\tan \theta_H \simeq \frac{{\sigma}_{xy}^{\rm int}}{{\sigma}_{xx}^{\rm Stokes}}+\frac{\eta^o}{\eta^e} \simeq \frac{l_{ee}}{R}\frac{\sigma_{xy}^{\rm int}}{{\sigma}_{xx}^{\rm Drude}}+\frac{\eta^o}{\eta^e}\,.
\end{equation}
For a generic value of $\sigma_{xy}^{\rm int}$, one expects to see a decrease of the Hall angle in the viscous regime compared to the elastic one. In the limit of zero
intrinsic conductivity, the Hall angle is determined by the odd and even viscosity ratio. While the value of the even viscosity is known, the accurate microscopic computations of the odd viscosity are yet to be done. 
However, based on the analogy of many-body skew scattering processes that are key for odd viscosity \cite{Messica2024} and impurity skew scattering \cite{Sinitsyn2007, Nagaosa2010, Konig2021}, we can estimate $\eta^o/\eta^e \simeq 1/(\epsilon_F \tau_{\rm{ee}})$.
From the experimental perspective, provided an independent measure of $\eta^e$ (see, for example, recent experiments in graphene \cite{Bandurin2018,Berdyugin2019,Waissman2024,Dean2024}), measurement of the Hall conductivity  in the hydrodynamic regime opens the possibility of extracting the odd viscosity $\eta^o$.

Finally, we note that a similar analysis extends beyond WSMs to other three-dimensional materials exhibiting both intrinsic and extrinsic mechanisms of Hall conductivity along with odd viscosity. The precise computation of the conductivities and odd viscosity is model-dependent.
However, once these quantities are computed for a specific model, we expect the behavior for other systems to be similar to the one studied in detail here. 
In particular, the transition between the elastic and viscous regimes and the corresponding change of the Hall angle are expected to hold. 

\textit{Conclusion}. We have studied the transition between the elastic and viscous hydrodynamic regimes in time-reversal broken Weyl semimetals.
Hall transport in these materials occurs through two parallel channels: one carried by the Fermi sea, which is robust against electron-electron collisions, and another carried by the Fermi surface. The latter is affected by the transition between the elastic and viscous regimes.

We carefully analyzed the flow around a sphere in an electronic fluid in the presence of odd viscosity and computed the conductivity tensors for a model random array of rare, opaque, and large spherical scatterers whose dimensions significantly exceed the electron wavelength.

We found qualitative changes in both longitudinal and transverse conductivities as the system transitions from the elastic to the viscous regime, occurring when the electron collision length becomes smaller than the impurity size. In the generic case of finite intrinsic conductivity, the Hall angle in the viscous regime is parametrically suppressed compared to the elastic regime. In the limiting case of zero intrinsic conductivity, the ratio of transverse to longitudinal conductivity equals the ratio of the odd and even components of the viscosity tensor.

\textit{Acknowledgments}. We are grateful to Polina Matveeva and the late Assa Auerbach for valuable discussions. We especially thank Igor Gornyi for his many critical and insightful comments on the final version of the manuscript.
 This work was supported by the National Science Foundation Grant No. DMR-2452658 and the H. I. Romnes Faculty Fellowship provided by the University of Wisconsin-Madison Office of the Vice Chancellor for Research and Graduate Education with funding from the Wisconsin Alumni Research Foundation (A. L.). Y. M. thanks the Ph.D. scholarship of the Israeli Scholarship Education Foundation (ISEF) for excellence in academic and social leadership.
 
\textit{Data availability}. The data that support the findings of this article are openly available \cite{StokesMathematicaData}.

\appendix

\section{Solution for the velocity field\label{sec:Appendix-A}}

To compute the  velocity profile we utilize perturbation theory, approximating Eq. (9) of the main text by  
\label{velocity_correction}
\begin{equation}
\label{NSlinearized_curlper}
\Delta\left(\nabla \times {\textbf v}_0+\gamma \nabla \times {\textbf v}_1+\frac{1}{2} \gamma \partial_z {\textbf v}_0\right) = 0 \,.
\end{equation}
The first term corresponds to the standard Stokes solution, Eq. (11) of the main text, and therefore identically vanishes. The last term can be easily computed
\begin{eqnarray}&&
\label{source}
\partial_z {\textbf v}_0=-\frac{Ru}{4}\left({\textbf A}_r\hat{r}+{\textbf A}_\theta\hat{\theta}+{\textbf A}_\phi\hat{\phi}\right),\\&&
{\textbf A}_r=\frac{9}{2}\sin2\theta\cos\phi\left(\frac{R^2}{r^4}-\frac{1}{r^2}\right),\\&&
{\textbf A}_\theta=-3\cos\phi\left(\frac{1}{r^2}+\frac{R^2}{r^4}\cos2\theta\right),\\&&
{\textbf A}_\phi=3\cos\theta\sin\phi\left(\frac{1}{r^2}+\frac{R^2}{r^4}\right).
\end{eqnarray} 
We look for a solution of this equation by expanding  
\begin{equation}
\label{Ansatz}
{\textbf v}_1(r,\theta,\phi)=\frac{u}{8}\sum_{n=0}^\infty\frac{R^{n}}{r^n}{\textbf V}_n(\theta,\phi)\,.
\end{equation}
Plugging  the ansatz Eq.~\eqref{Ansatz} together with  Eq.~\eqref{source} into  Eq.~\eqref{NSlinearized_curlper} one finds
\begin{eqnarray}&&
\label{particular}
{\textbf v}_1(r,\theta,\phi)=-\frac{uR}{8}\bigg[\frac{3}{r}\sin\theta\sin\phi \hat{\mathbf{r}}+\frac{3R^2}{2r^3}\cos\theta\sin\phi\hat{\bm{\theta}}+\nonumber \\&&
+\cos\phi\left(\frac{3}{2r}(1-\cos2\theta)+\frac{3R^2}{2r^3}\cos2\theta\right)\hat{\bm{\phi}}\bigg].
\end{eqnarray}
To satisfy the boundary conditions ${\textbf v}_1(R,\theta,\phi)=0$ one needs to compute the zero modes of the operator $\Delta {\rm curl}$, and add them to the particular solution of Eq. (\ref{particular}). 
Because the velocity field is divergence-free, the zero modes can be constructed as a curl of a vector field.
It is convenient to use the vector spherical harmonics, defined as
\begin{eqnarray}&&
{\textbf Y}_{lm}^1(\theta,\phi)=Y_{lm}(\theta,\phi)\hat{\mathbf{r}},\nonumber \\&&
{\textbf Y}_{lm}^2(\theta,\phi)=\frac{\partial Y_{lm}(\theta,\phi)}{\partial \theta}\hat{\bm{\theta}}
+\frac{1}{\sin \theta} \frac{\partial Y_{lm}(\theta,\phi)}{\partial \phi}\hat{\bm{\phi}},
\nonumber \\&&
{\textbf Y}_{lm}^3(\theta,\phi)=\frac{1}{\sin\theta}\frac{\partial Y_{lm}(\theta,\phi)}{\partial \phi}\hat{\bm{\theta}}
-\frac{\partial Y_{lm}(\theta,\phi)}{\partial \theta}\hat{\bm{\phi}}.
\end{eqnarray}
We need the sub-space of the  zero modes of the vector Laplace operator, spanned by three vectors  
\begin{eqnarray}&&
{\textbf Z}_1=\nabla\times\left(\frac{Y_1^0}{r^2}\hat{\mathbf{x}}\right),\nonumber\\&&
{\textbf Z}_2=\nabla\times\left(\frac{Y_1^{1}-Y_1^{-1}}{r^2}\hat{\mathbf{z}}\right),\nonumber \\&&
{\textbf Z}_3=\nabla\times\left(\frac{{\textbf Y}_{1,1}^{3}+{\textbf Y}^3_{1,-1}}{r^2}\right).
\end{eqnarray}
In addition, there is  
\begin{eqnarray}
{\textbf Z}_4=\nabla\times\left({\textbf Y}_{1,1}^{3}+{\textbf Y}_{1,-1}^{3}\right),
\end{eqnarray}
that is a zero mode of the full operator $\Delta \nabla\times {\textbf Z}_4=0$, even though 
$\Delta{\textbf Z}_4 \neq 0$. 
The boundary condition for the velocity to vanish at infinity is automatically satisfied. 
Imposing the condition ${\textbf v}_1(R,\theta,\phi)=0$ one arrives at Eq. (12) of the main text.

\section{The odd part of the viscosity tensor and calculation of the stress forces\label{sec:Appendix-B}}

\subsection{Components of the odd viscous tensor}

The components of the odd part of the stress tensor for the odd viscosity model [Eq. (5) of the main text] in the Cartesian coordinates are 

\begin{subequations}
	\begin{align}
	&\sigma^o_{11}=\eta^o\left(\frac{\partial v_1}{\partial x_2}+\frac{\partial v_2}{\partial x_1}\right), \\
	&\sigma^o_{12}=\eta^o\left(\frac{\partial v_2}{\partial x_2}-\frac{\partial v_1}{\partial x_1}\right), \\
	&\sigma^o_{13}=\frac{1}{2} \eta^0\left(\frac{\partial v_2}{\partial x_3}+\frac{\partial v_3}{\partial x_2}\right),\\
	&\sigma^o_{22}=-\eta^o\left(\frac{\partial v_1}{\partial x_2}+\frac{\partial v_2}{\partial x_1}\right), \\
	&\sigma^o_{23}=-\frac{1}{2} \eta^o\left(\frac{\partial v_1}{\partial x_3}+\frac{\partial v_3}{\partial x_1}\right),\\
	&\sigma^o_{33}=0\,,
	\end{align}
\end{subequations}
where the symmetry of the tensor determines the remaining components.
Using this tensor in the spherical coordinates is convenient,  denoted as  $\tilde{\sigma}^o$. 
To find the tensor in the spherical coordinates one uses the standard  transformation matrix, 
\begin{subequations}
	\begin{equation}
	\begin{bmatrix}
	\hat{\mathbf{x}} \\
	\hat{\mathbf{y}} \\
	\hat{\mathbf{z}}
	\end{bmatrix}
	=
	T
	\begin{bmatrix}
	\hat{\mathbf{r}} \\
	\hat{\bm{\theta}} \\
	\hat{\bm{\phi}}
	\end{bmatrix},
	\end{equation}
	where
	\begin{equation}
	T =
	\begin{bmatrix}
	\sin\theta \cos\phi & \cos\theta \cos\phi & -\sin\phi \\
	\sin\theta \sin\phi & \cos\theta \sin\phi & \cos\phi \\
	\cos\theta & -\sin\theta & 0
	\end{bmatrix}\,.
	\end{equation}
\end{subequations}
The tensor in the spherical coordinates is thus given by 
\begin{equation}
\tilde{\sigma}_{mn}=\sum_{m_1,m_2=1}^3T_{m_1,m}T_{m_2,n}\sigma_{m_1,m_2}\,.
\end{equation}
The transformation of the derivatives  is governed by 
\begin{subequations}
	\begin{equation}    
	\begin{bmatrix}
	\partial_x \\
	\partial_y \\
	\partial_z
	\end{bmatrix}
	=
	A
	\begin{bmatrix}
	\partial_r \\
	\partial_\theta \\
	\partial_\phi
	\end{bmatrix}, 
	\end{equation}
	where
	\begin{equation}
	A=
	\begin{bmatrix}
	\sin\theta \cos\phi & \frac{\cos\theta \cos\phi}{r} & -\frac{\sin\phi}{r\sin \theta} \\
	\sin\theta \sin\phi & \frac{\cos\theta \sin\phi}{r} & \frac{\cos\phi}{r \sin \theta} \\
	\cos\theta & -\frac{\sin\theta}{r} & 0
	\end{bmatrix}\,.
	\end{equation}
\end{subequations}
We can now express the components of the stress tensor 
in Cartesian coordinates in terms of the velocity and its derivatives computed in spherical coordinates, denoted by  tildes,
\begin{eqnarray} &&
\label{sigmao}
\sigma^o_{11}=\eta^0(A_{2,i_1}\tilde{\partial}_{i_1}T_{1,k}\tilde{v}_k+A_{1,i_1}\tilde{\partial}_{i_1}T_{2,k}\tilde{v}_k),\nonumber \\&&
\sigma^o_{12}=\eta^0(A_{2,i_1}\tilde{\partial}_{i_1}T_{2,k}\tilde{v}_k+A_{2,i_1}\tilde{\partial}_{i_1}T_{1,k}\tilde{v}_k),\nonumber \\&&
\sigma^o_{13}=\frac{1}{2} \eta^0(A_{3,i_1}\tilde{\partial}_{i_1}T_{2,k}\tilde{v}_k+A_{2,i_1}\tilde{\partial}_{i_1}T_{3,k}\tilde{v}_k),\nonumber \\&&
\sigma^o_{23}=-\frac{1}{2} \eta^0(A_{3,i_1}\tilde{\partial}_{i_1}T_{1,k}\tilde{v}_k+A_{1,i_1}\tilde{\partial}_{i_1}T_{3,k}\tilde{v}_k),\nonumber \\&&
\sigma^o_{22}=-\sigma^o_{11}\,.
\end{eqnarray}
The full expressions are too complicated to be written explicitly but can be easily handled in Mathematica. 

\subsection{Forces from the viscous stress tensors}

The components of the even part of the viscous stress tensor are well known \cite{Landau2003}.
To compute forces we need the following components
\begin{subequations}
	\begin{align}
	&\sigma^e_{rr}(v)=2\eta^e \frac{\partial v_r}{\partial r}, \\
	&\sigma_{r,\theta}(v)=\eta^e\left(\frac{1}{r}\frac{\partial v_r}{\partial \theta}+
	\frac{\partial v_\theta}{\partial r}
	-\frac{v_\theta}{ r}\right),  \\
	&\sigma_{r,\phi}(v)=\eta^e\left(\frac{\partial v_\phi}{\partial r}+
	\frac{1}{r\sin \theta} \frac{\partial v_r}{\partial \phi}
	-\frac{v_\phi}{ r}\right). 
	\end{align}
\end{subequations}
Computing the components of the tensor $\sigma^e(\gamma v_1)$ at the surface of the sphere one finds
\begin{subequations}
	\begin{align}
	&\sigma_{rr}^e(\gamma v_1)=0 ,  \\
	&\sigma_{r\theta}^e(\gamma v_1)=\frac{3\eta^ou}{8R}\cos\theta\sin\phi \,,\\
	&\sigma_{r,\phi}^e(\gamma v_1)=\frac{3\eta^ou}{8R}\cos\phi \cos 2\theta\, .
	\end{align}
\end{subequations}
The resulting force per unit area is given by
\begin{eqnarray}&&
F_x^{I}(\theta,\phi)=\frac{3\eta^o}{8R}\left(\cos^2\theta\sin\phi\cos\phi-
\cos2\theta\cos\phi\sin\phi\right)\,, \nonumber\\&&
F_y^{I}(\theta,\phi)=\frac{3\eta^o}{8R}\left(\cos^2\theta\sin^2\phi +\cos2\theta
\cos^2\phi\right) \,.
\end{eqnarray}
Integrating over the surface of the sphere, one finds
\begin{eqnarray}&&
F^{\sigma-I}_x=R^2\int_0^\pi\sin\theta d\theta \int_0^{2\pi}d\phi F^{I}_x(\theta,\phi)=0\,, \nonumber\\&&
F^{\sigma-I}_y=R^2\int_0^\pi\sin\theta d\theta \int_0^{2\pi}d\phi F^{I}_y(\theta,\phi)=0   \,.
\end{eqnarray}
Next, we compute the force density coming from  $\sigma^o(v_0)$, Eq. (\ref{sigmao}),
\begin{align}
F^{II}_y(\theta,\phi)=\frac{3\eta^o u}{4 R} (\sin^2\theta \cos 2\phi -1).
\end{align}
The integrated force in the transverse direction is given by
\begin{equation}
F^{\sigma-II}_y=R^2\int_0^\pi\sin\theta d\theta\int_0^{2\pi}d\phi F^{II}_y(\theta,\phi)
=-3\pi uR\eta^o.
\end{equation}
On the $x$ direction, the integrated force vanishes. This force has to be added to the corresponding contribution arising from the pressure [Eq. (15) of the main text], which comes with a different numerical coefficient $3\pi/2$ and a positive sign. As a result, the total Hall component of the force adds to the result given in Eq. (18) of the main text. 

\section{Anomalous Hall conductivity analysis\label{sec:Appendix-C}}
\subsection{Intrinsic and extrinsic mechanisms of transport}

Here we give the details of the analysis done in the second half of the paper for the calculation of the conductivity in the hydrodynamic regime. In Weyl semimetals, the current can be separated into an intrinsic
and an extrinsic part. The intrinsic part is carried by the Fermi-sea
electrons and emerges due to the Berry curvature-induced velocity
in the presence of an electric field. On the linear response level, this
effect does not require any change of occupation of the electronic states, and one may approximate the distribution function as an equilibrium one. This is why this channel
is robust against scattering, both elastic and inelastic.

The extrinsic part of the current is carried by electrons at the Fermi surface. This
contribution is finite only when the electrons have a non-equilibrium distribution
in the phase space. Therefore, the extrinsic part of the conductivity is
sensitive to the 'shift' of the Fermi surface. This is why scattering
processes affect this channel. With this logic in mind, the electric
current is given by a sum of the currents carried by the intrinsic and the extrinsic channels,
\begin{equation}
\textbf{j}=\textbf{j}_{\textrm{int}}+\textbf{j}_{\textrm{ext}}.\label{eq:j_total}
\end{equation}
The extrinsic current can be expressed in terms of a single particle
distribution function in the standard way,
\begin{equation}
\label{eq:j_ext}
\textbf{j}_{\textrm{ext}}=e\intop \left( \textrm{d} \textbf{k} \right) f(\textbf{k})\textbf{v}_{\textbf{k}},
\end{equation}
with $e$ denoting the electron charge. Note that due to side jumps the velocity operator in Eq. (\ref{eq:j_ext})
acquires an additional term \cite{Sinitsyn2006, Atencia2022}. It can be interpreted as a coordinate shift in the electron's trajectory after the scattering event. The computation of the
side-jump part of the velocity operator is model dependent and can be
found in Ref. \cite{Sinitsyn2007}. While the modification of
the velocity operator plays a role in non-interacting WSMs, these terms
are simplified in the hydrodynamic limit.

The intrinsic current is given in terms of Berry curvature integrated
over the filled states \cite{Sinitsyn2007}
\begin{equation}
\label{eq:j_int}
\textbf{j}_{\textrm{int}}=e\intop \left( \textrm{d} \textbf{k} \right) f_{0}(\epsilon_{\textbf{k}})e\textbf{E}\times {\bf{\Omega}}_{\textbf{k}}.
\end{equation}
Here we approximate the distribution
function by the equilibrium one. Equations (\ref{eq:j_total})-(\ref{eq:j_int}) are general and hold for all types of electron-disorder
and electron-electron scattering processes (with the details of the
side-jump processes encoded in the velocity operator $\textbf{v}_{\textbf{k}}$).
We next consider the transition to the hydrodynamic limit, assuming that the electron-electron mean free path $l_{\textrm{ee}}$ is much smaller than the electron-disorder mean free path $l_{\textrm{imp}}$. In this limit, we separate
the zero modes of the e-e collision integral from the finite modes. The zero modes correspond to the conserved hydrodynamic
quantities, and the finite modes give rise to the viscosity and other
dissipative coefficients \cite{Landau2003}. The expression
for the extrinsic current reduces to the product of the particle density
and the hydrodynamic velocity $\textbf{u}$,
\begin{equation}
\textbf{j}_{\textrm{ext}}=en\textbf{u}.
\end{equation}
It is worth stressing that the intrinsic part of the current [Eq.
(\ref{eq:j_int})] is unaffected by the transition to the hydrodynamic
regime.

We now outline the computation of the conductivity in the hydrodynamic
regime. By multiplying the Boltzmann equation by the momentum $\textbf{k}$, integrating
over $\textbf{k}$, and neglecting nonlinear terms in the hydrodynamic velocity
${\textbf u}$, we arrive at the force balance equation 
\begin{equation}
\frac{w}{v_{F}^{2}}\partial_{t}u_{i}=enE_{i}+\partial_{j}\sigma_{ij}+\int\left(\textrm{d} \textbf{k}\right)k_{i}I_{\textrm{imp}}\left[f_{\textbf{k}}\right], \label{eq:force balance}
\end{equation}
where $w$ is the enthalpy density, $n$ is the electron particle density, $I_{\textrm{imp}}\left[f\right]$
is the electron-disorder collision integral, and $\sigma_{ij}$
is the stress tensor given by \footnote{In this section, we denote the entire stress tensor (both the pressure and viscous parts) by $\sigma_{ij}$. In the main text, we explicitly write the pressure term in the Navier-Stokes equation [Eq. (1) of the main text], and $\sigma_{ij}$ there denotes only the viscous part of the stress tensor.} \cite{Narozhny2019}

\begin{equation}
\sigma_{ij}=\intop\left(\textrm{d} \textbf{k}\right)k_{i}v_{j}(\textbf{k})f(\textbf{k}).
\end{equation}
The stress tensor includes a part corresponding to the pressure, which originates from the zero-modes part of the distribution function, and a viscous part coming from the finite modes. Hence, for a precise calculation of the viscous part one needs to solve the distribution function, see for example Ref. \cite{Muller2009}. Based on the symmetries of the
system, the viscosity tensor has two independent components corresponding
to the even (shear) and odd (Hall) viscosities, Eqs. (2) and (3) in the main
text. Let us note that a similar equation to Eq.
(\ref{eq:force balance}) for the electron fluid in a WSM in the presence
of a magnetic field and non-linear terms in ${\textbf u}$ was recently
derived in Ref. \cite{Gorbar2018}.

\subsection{Impurity-dominated and viscous-dominated regimes}

To proceed with the analysis of the electric conductivity, the details of the disorder need to be specified. We model the disorder as made of finite-sized impurities, represented by spherical hard-wall
potentials of radius $R$, put at low concentration $n_{\textrm{imp}} \ll 1/R^3$. We mention that large scatterers were found to have universal properties irrespective of their particular shape \cite{Yudson2007}, and thus we expect the results obtained here to remain qualitatively applicable for other types of impurities. The corresponding electron-disorder mean free path
is given by $l_{\textrm{imp}}=1/\left(n_{{\rm imp}}R^{2}\right)$. We analyze the behavior in the hydrodynamic regime in the two limits
$R\ll l_{\textrm{ee}}$ and $l_{\textrm{ee}}\ll R$. 

We start with the limit $R\ll l_{\textrm{ee}}\ll l_{\textrm{imp}}$. In this case,
the momentum loss due to impurity scattering  can be estimated as
\begin{equation}
\int\left(\textrm{d} \textbf{k}\right)k_{i}I_{\textrm{imp}}\left[f_{\textbf{k}}\right]= -\frac{w}{v_{F}^{2}\tau_{\textrm{imp}}}u_{i},
\end{equation}
with $\tau_{\textrm{imp}}=l_{\textrm{imp}}/v_F$. 
In this limit, the viscous part of the stress tensor can be neglected, 
and the boost velocity does not depend on $l_{ee}$. Consequently, the electric conductivity is governed by electron-disorder scattering. Thus, for $R\ll l_{ee}$, the transition to the hydrodynamic regime ($l_{ee}\ll l_{imp}$) does not change the electric conductivity, apart from possibly modifying numerical prefactors \cite{gantmakher1987carrier}.

The reason that the electric conductivity is robust against e-e scattering is due to the fact that e-e collisions conserve the total electronic momentum, which is equivalent to the charge current for a parabolic spectrum \cite{Principi2015} (an appropriate approximation for the WSM in the Fermi-liquid regime, $\mu/T\gg1$). This argument therefore relies on the chemical potential being far from the neutrality point, as consistently assumed in this paper. 

In the second limit, $l_{\textrm{ee}}\ll R\ll l_{\textrm{\textrm{imp}}}$,
which is the focus of this work, the viscosity has an important effect
in the vicinity of the impurities. The disorder scattering term in
Eq. (\ref{eq:force balance}) imposes  the no-slip
boundary conditions for the boost velocity on the surface of the impurities \cite{Kiselev2019} \footnote{More generally, one may define the slip length $\xi$ which generalizes the
	no-slip ($\xi=0$) and no-stress ($\xi=\infty$) boundary conditions, see Refs. \cite{Kiselev2019, Moessner2019}. In 3D Fermi liquids, the slip length scale is in the order of $l_{\textrm{ee}}$ \cite{Kiselev2019}. As $l_{\textrm{ee}}$ is the smallest length scale in our case, it is safe to take the no-slip boundary condition.}.
The problem is thus equivalent to the flow of a viscous liquid through randomly distributed finite-sized obstacles. The electronic momentum is lost through the viscous forces between the liquid and the obstacles \footnote{For a comprehensive study in a more general range of the length scales
	$l_{\textrm{ee}}$, $R$, and $l_{\textrm{\textrm{imp}}}$, we refer
	to Ref. \cite{Guo2016}. The electron-electron scattering
	can be interpreted as causing a renormalization of the impurity scattering rate, as
	the emergent collective motion of the electrons leads them to tend
	to avoid disorder scattering. In the extreme hydrodynamic limit where
	$l_{\rm{ee}}\ll R$, all the electronic momentum is lost from electrons
	that collide with an impurity and then quickly (after travelling
	a short distance $l_{\textrm{ee}}$) equilibrate with the bulk hydrodynamic
	flow, thus forming a thin boundary layer near the impurity.}.
Thus, the momentum dissipation rate can be calculated by integrating
the stress tensor over an impurity boundary. Multiplying by the
impurity concentration $n_\textrm{imp}$, one obtains the average force density acting on the liquid. In the steady state, the viscous force is balanced by the electric field,
\begin{equation}
\textbf{E}=\frac{1}{en}n_{\textrm{imp}}\textbf{F}.
\end{equation}
The force  ${\textbf F}$ acting 
on a single impurity by the fluid that flows around it with relative velocity ${\textbf u}$ was computed in Eqs. (17) and (18) of the main text (with ${\textbf F}={\textbf F}_\parallel + {\textbf F}_{\textrm{Hall}}$).
It can be written as 
\begin{equation}
\textbf{F} = \zeta\textbf{u},
\end{equation}
with $\zeta$ being the drag tensor
\begin{equation}
\zeta \equiv \pi\eta^{\textrm{e}} R\left(
\begin{array}{cc}
6 & -\frac{3}{2}\frac{\eta^{\textrm{o}}}{\eta^{\textrm{e}}} \\
\frac{3}{2}\frac{\eta^{\textrm{o}}}{\eta^{\textrm{e}}} & 6
\end{array}\right)\,.
\end{equation}
The off-diagonal components of $\zeta$ are antisymmetric, as expected for Hall-like responses.

We then calculate the external conductivities $\sigma_{\alpha\beta}^{\textrm{\textrm{ext}}}\equiv j_{\alpha}^{\textrm{ext}}/E_{\beta}=enu_{\alpha}/E_{\beta}$
and find

\begin{eqnarray}
&  & \sigma_{xx}^{\textrm{ext}}=\left(en\right)^{2}\frac{6}{\pi n_{\textrm{imp}}\eta^{\textrm{e}}R\left(36+\frac{9\eta^{\textrm{o}}}{4\eta^{\textrm{e}}}\right)},\\
&  & \sigma_{xy}^{\textrm{ext}}=\frac{\eta^{\textrm{o}}}{4\eta^{\textrm{e}}}\sigma_{xx}^{\textrm{ext}}.
\end{eqnarray}
Having obtained the ratio of the extrinsic conductivities $\sigma_{xy}^{\textrm{ext}}/\sigma_{xx}^{\textrm{ext}}\simeq \eta^{\textrm{o}}/\eta^{\textrm{e}}$, we arrive at the scaling of the anomalous Hall conductivity in the viscous-dominated regime, Eq. (23) of the main text.

\bibliography{stokes_bib}

\begin{thebibliography}{69}%
\makeatletter
\providecommand \@ifxundefined [1]{%
 \@ifx{#1\undefined}
}%
\providecommand \@ifnum [1]{%
 \ifnum #1\expandafter \@firstoftwo
 \else \expandafter \@secondoftwo
 \fi
}%
\providecommand \@ifx [1]{%
 \ifx #1\expandafter \@firstoftwo
 \else \expandafter \@secondoftwo
 \fi
}%
\providecommand \natexlab [1]{#1}%
\providecommand \enquote  [1]{``#1''}%
\providecommand \bibnamefont  [1]{#1}%
\providecommand \bibfnamefont [1]{#1}%
\providecommand \citenamefont [1]{#1}%
\providecommand \href@noop [0]{\@secondoftwo}%
\providecommand \href [0]{\begingroup \@sanitize@url \@href}%
\providecommand \@href[1]{\@@startlink{#1}\@@href}%
\providecommand \@@href[1]{\endgroup#1\@@endlink}%
\providecommand \@sanitize@url [0]{\catcode `\\12\catcode `\$12\catcode
  `\&12\catcode `\#12\catcode `\^12\catcode `\_12\catcode `\%12\relax}%
\providecommand \@@startlink[1]{}%
\providecommand \@@endlink[0]{}%
\providecommand \url  [0]{\begingroup\@sanitize@url \@url }%
\providecommand \@url [1]{\endgroup\@href {#1}{\urlprefix }}%
\providecommand \urlprefix  [0]{URL }%
\providecommand \Eprint [0]{\href }%
\providecommand \doibase [0]{https://doi.org/}%
\providecommand \selectlanguage [0]{\@gobble}%
\providecommand \bibinfo  [0]{\@secondoftwo}%
\providecommand \bibfield  [0]{\@secondoftwo}%
\providecommand \translation [1]{[#1]}%
\providecommand \BibitemOpen [0]{}%
\providecommand \bibitemStop [0]{}%
\providecommand \bibitemNoStop [0]{.\EOS\space}%
\providecommand \EOS [0]{\spacefactor3000\relax}%
\providecommand \BibitemShut  [1]{\csname bibitem#1\endcsname}%
\let\auto@bib@innerbib\@empty
\bibitem [{\citenamefont {Bandurin}\ \emph {et~al.}(2016)\citenamefont
  {Bandurin}, \citenamefont {Torre}, \citenamefont {Kumar}, \citenamefont
  {Shalom}, \citenamefont {Tomadin}, \citenamefont {Principi}, \citenamefont
  {Auton}, \citenamefont {Khestanova}, \citenamefont {Novoselov}, \citenamefont
  {Grigorieva}, \citenamefont {Ponomarenko}, \citenamefont {Geim},\ and\
  \citenamefont {Polini}}]{Bandurin2016}%
  \BibitemOpen
  \bibfield  {author} {\bibinfo {author} {\bibfnamefont {D.~A.}\ \bibnamefont
  {Bandurin}}, \bibinfo {author} {\bibfnamefont {I.}~\bibnamefont {Torre}},
  \bibinfo {author} {\bibfnamefont {R.~K.}\ \bibnamefont {Kumar}}, \bibinfo
  {author} {\bibfnamefont {M.~B.}\ \bibnamefont {Shalom}}, \bibinfo {author}
  {\bibfnamefont {A.}~\bibnamefont {Tomadin}}, \bibinfo {author} {\bibfnamefont
  {A.}~\bibnamefont {Principi}}, \bibinfo {author} {\bibfnamefont {G.~H.}\
  \bibnamefont {Auton}}, \bibinfo {author} {\bibfnamefont {E.}~\bibnamefont
  {Khestanova}}, \bibinfo {author} {\bibfnamefont {K.~S.}\ \bibnamefont
  {Novoselov}}, \bibinfo {author} {\bibfnamefont {I.~V.}\ \bibnamefont
  {Grigorieva}}, \bibinfo {author} {\bibfnamefont {L.~A.}\ \bibnamefont
  {Ponomarenko}}, \bibinfo {author} {\bibfnamefont {A.~K.}\ \bibnamefont
  {Geim}},\ and\ \bibinfo {author} {\bibfnamefont {M.}~\bibnamefont {Polini}},\
  }\bibfield  {title} {\bibinfo {title} {Negative local resistance caused by
  viscous electron backflow in graphene},\ }\href
  {https://doi.org/10.1126/science.aad0201} {\bibfield  {journal} {\bibinfo
  {journal} {Science}\ }\textbf {\bibinfo {volume} {351}},\ \bibinfo {pages}
  {1055} (\bibinfo {year} {2016})}\BibitemShut {NoStop}%
\bibitem [{\citenamefont {Kumar}\ \emph {et~al.}(2017)\citenamefont {Kumar},
  \citenamefont {Bandurin}, \citenamefont {Pellegrino}, \citenamefont {Cao},
  \citenamefont {Principi}, \citenamefont {Guo}, \citenamefont {Auton},
  \citenamefont {Shalom}, \citenamefont {Ponomarenko}, \citenamefont
  {Falkovich}, \citenamefont {Watanabe}, \citenamefont {Taniguchi},
  \citenamefont {Grigorieva}, \citenamefont {Levitov}, \citenamefont {Polini},\
  and\ \citenamefont {Geim}}]{Kumar2017}%
  \BibitemOpen
  \bibfield  {author} {\bibinfo {author} {\bibfnamefont {R.~K.}\ \bibnamefont
  {Kumar}}, \bibinfo {author} {\bibfnamefont {D.~A.}\ \bibnamefont {Bandurin}},
  \bibinfo {author} {\bibfnamefont {F.~M.}\ \bibnamefont {Pellegrino}},
  \bibinfo {author} {\bibfnamefont {Y.}~\bibnamefont {Cao}}, \bibinfo {author}
  {\bibfnamefont {A.}~\bibnamefont {Principi}}, \bibinfo {author}
  {\bibfnamefont {H.}~\bibnamefont {Guo}}, \bibinfo {author} {\bibfnamefont
  {G.~H.}\ \bibnamefont {Auton}}, \bibinfo {author} {\bibfnamefont {M.~B.}\
  \bibnamefont {Shalom}}, \bibinfo {author} {\bibfnamefont {L.~A.}\
  \bibnamefont {Ponomarenko}}, \bibinfo {author} {\bibfnamefont
  {G.}~\bibnamefont {Falkovich}}, \bibinfo {author} {\bibfnamefont
  {K.}~\bibnamefont {Watanabe}}, \bibinfo {author} {\bibfnamefont
  {T.}~\bibnamefont {Taniguchi}}, \bibinfo {author} {\bibfnamefont {I.~V.}\
  \bibnamefont {Grigorieva}}, \bibinfo {author} {\bibfnamefont {L.~S.}\
  \bibnamefont {Levitov}}, \bibinfo {author} {\bibfnamefont {M.}~\bibnamefont
  {Polini}},\ and\ \bibinfo {author} {\bibfnamefont {A.~K.}\ \bibnamefont
  {Geim}},\ }\bibfield  {title} {\bibinfo {title} {Superballistic flow of
  viscous electron fluid through graphene constrictions},\ }\href
  {https://doi.org/10.1038/nphys4240} {\bibfield  {journal} {\bibinfo
  {journal} {Nat. Phys.}\ }\textbf {\bibinfo {volume} {13}},\ \bibinfo {pages}
  {1182} (\bibinfo {year} {2017})}\BibitemShut {NoStop}%
\bibitem [{\citenamefont {Xiao}\ \emph {et~al.}(2010)\citenamefont {Xiao},
  \citenamefont {Chang},\ and\ \citenamefont {Niu}}]{Xiao2010}%
  \BibitemOpen
  \bibfield  {author} {\bibinfo {author} {\bibfnamefont {D.}~\bibnamefont
  {Xiao}}, \bibinfo {author} {\bibfnamefont {M.~C.}\ \bibnamefont {Chang}},\
  and\ \bibinfo {author} {\bibfnamefont {Q.}~\bibnamefont {Niu}},\ }\bibfield
  {title} {\bibinfo {title} {Berry phase effects on electronic properties},\
  }\href {https://doi.org/10.1103/RevModPhys.82.1959} {\bibfield  {journal}
  {\bibinfo  {journal} {RMP}\ }\textbf {\bibinfo {volume} {82}},\ \bibinfo
  {pages} {1959} (\bibinfo {year} {2010})}\BibitemShut {NoStop}%
\bibitem [{\citenamefont {Avron}\ \emph {et~al.}(1995)\citenamefont {Avron},
  \citenamefont {Seiler},\ and\ \citenamefont {Zograf}}]{Avron1995}%
  \BibitemOpen
  \bibfield  {author} {\bibinfo {author} {\bibfnamefont {J.~E.}\ \bibnamefont
  {Avron}}, \bibinfo {author} {\bibfnamefont {R.}~\bibnamefont {Seiler}},\ and\
  \bibinfo {author} {\bibfnamefont {P.~G.}\ \bibnamefont {Zograf}},\ }\bibfield
   {title} {\bibinfo {title} {Viscosity of quantum {Hall} fluids},\ }\href
  {https://doi.org/10.1103/PhysRevLett.75.697} {\bibfield  {journal} {\bibinfo
  {journal} {Phys. Rev. Lett.}\ }\textbf {\bibinfo {volume} {75}},\ \bibinfo
  {pages} {697} (\bibinfo {year} {1995})}\BibitemShut {NoStop}%
\bibitem [{\citenamefont {Avron}(1998)}]{Avron1998}%
  \BibitemOpen
  \bibfield  {author} {\bibinfo {author} {\bibfnamefont {J.~E.}\ \bibnamefont
  {Avron}},\ }\bibfield  {title} {\bibinfo {title} {Odd viscosity},\ }\href
  {https://doi.org/10.1023/a:1023084404080} {\bibfield  {journal} {\bibinfo
  {journal} {J. Stat. Phys.}\ }\textbf {\bibinfo {volume} {92}},\ \bibinfo
  {pages} {543} (\bibinfo {year} {1998})}\BibitemShut {NoStop}%
\bibitem [{\citenamefont {Fruchart}\ \emph {et~al.}(2023)\citenamefont
  {Fruchart}, \citenamefont {Scheibner},\ and\ \citenamefont
  {Vitelli}}]{Fruchart2023}%
  \BibitemOpen
  \bibfield  {author} {\bibinfo {author} {\bibfnamefont {M.}~\bibnamefont
  {Fruchart}}, \bibinfo {author} {\bibfnamefont {C.}~\bibnamefont
  {Scheibner}},\ and\ \bibinfo {author} {\bibfnamefont {V.}~\bibnamefont
  {Vitelli}},\ }\bibfield  {title} {\bibinfo {title} {Odd viscosity and odd
  elasticity},\ }\href
  {https://doi.org/10.1146/annurev-conmatphys-040821-125506} {\bibfield
  {journal} {\bibinfo  {journal} {Annu. Rev. Condens. Matter Phys.}\ }\textbf
  {\bibinfo {volume} {14}},\ \bibinfo {pages} {471} (\bibinfo {year}
  {2023})}\BibitemShut {NoStop}%
\bibitem [{\citenamefont {Banerjee}\ \emph {et~al.}(2017)\citenamefont
  {Banerjee}, \citenamefont {Souslov}, \citenamefont {Abanov},\ and\
  \citenamefont {Vitelli}}]{Banerjee2017}%
  \BibitemOpen
  \bibfield  {author} {\bibinfo {author} {\bibfnamefont {D.}~\bibnamefont
  {Banerjee}}, \bibinfo {author} {\bibfnamefont {A.}~\bibnamefont {Souslov}},
  \bibinfo {author} {\bibfnamefont {A.~G.}\ \bibnamefont {Abanov}},\ and\
  \bibinfo {author} {\bibfnamefont {V.}~\bibnamefont {Vitelli}},\ }\bibfield
  {title} {\bibinfo {title} {Odd viscosity in chiral active fluids},\ }\href
  {https://doi.org/10.1038/s41467-017-01378-7} {\bibfield  {journal} {\bibinfo
  {journal} {Nat. Commun.}\ }\textbf {\bibinfo {volume} {8}},\ \bibinfo {pages}
  {1573} (\bibinfo {year} {2017})}\BibitemShut {NoStop}%
\bibitem [{\citenamefont {Markovich}\ and\ \citenamefont
  {Lubensky}(2021)}]{Markovich2021}%
  \BibitemOpen
  \bibfield  {author} {\bibinfo {author} {\bibfnamefont {T.}~\bibnamefont
  {Markovich}}\ and\ \bibinfo {author} {\bibfnamefont {T.~C.}\ \bibnamefont
  {Lubensky}},\ }\bibfield  {title} {\bibinfo {title} {Odd viscosity in active
  matter: Microscopic origin and {3D} effects},\ }\href
  {https://doi.org/10.1103/PhysRevLett.127.048001} {\bibfield  {journal}
  {\bibinfo  {journal} {Phys. Rev. Lett.}\ }\textbf {\bibinfo {volume} {127}},\
  \bibinfo {pages} {048001} (\bibinfo {year} {2021})}\BibitemShut {NoStop}%
\bibitem [{\citenamefont {Hosaka}\ \emph {et~al.}(2023)\citenamefont {Hosaka},
  \citenamefont {Golestanian},\ and\ \citenamefont {Vilfan}}]{Hosaka2023}%
  \BibitemOpen
  \bibfield  {author} {\bibinfo {author} {\bibfnamefont {Y.}~\bibnamefont
  {Hosaka}}, \bibinfo {author} {\bibfnamefont {R.}~\bibnamefont
  {Golestanian}},\ and\ \bibinfo {author} {\bibfnamefont {A.}~\bibnamefont
  {Vilfan}},\ }\bibfield  {title} {\bibinfo {title} {Lorentz reciprocal theorem
  in fluids with odd viscosity},\ }\href
  {https://doi.org/10.1103/PhysRevLett.131.178303} {\bibfield  {journal}
  {\bibinfo  {journal} {Phys. Rev. Lett.}\ }\textbf {\bibinfo {volume} {131}},\
  \bibinfo {pages} {178303} (\bibinfo {year} {2023})}\BibitemShut {NoStop}%
\bibitem [{\citenamefont {Lier}\ \emph {et~al.}(2023)\citenamefont {Lier},
  \citenamefont {Duclut}, \citenamefont {Bo}, \citenamefont {Armas},
  \citenamefont {J{\"u}licher},\ and\ \citenamefont {Sur{\'o}wka}}]{Lier2023}%
  \BibitemOpen
  \bibfield  {author} {\bibinfo {author} {\bibfnamefont {R.}~\bibnamefont
  {Lier}}, \bibinfo {author} {\bibfnamefont {C.}~\bibnamefont {Duclut}},
  \bibinfo {author} {\bibfnamefont {S.}~\bibnamefont {Bo}}, \bibinfo {author}
  {\bibfnamefont {J.}~\bibnamefont {Armas}}, \bibinfo {author} {\bibfnamefont
  {F.}~\bibnamefont {J{\"u}licher}},\ and\ \bibinfo {author} {\bibfnamefont
  {P.}~\bibnamefont {Sur{\'o}wka}},\ }\bibfield  {title} {\bibinfo {title}
  {Lift force in odd compressible fluids},\ }\href
  {https://doi.org/10.1103/PhysRevE.108.L023101} {\bibfield  {journal}
  {\bibinfo  {journal} {Phys. Rev. E}\ }\textbf {\bibinfo {volume} {108}},\
  \bibinfo {pages} {L023101} (\bibinfo {year} {2023})}\BibitemShut {NoStop}%
\bibitem [{\citenamefont {Everts}\ and\ \citenamefont
  {Cichocki}(2024)}]{Everts2024}%
  \BibitemOpen
  \bibfield  {author} {\bibinfo {author} {\bibfnamefont {J.~C.}\ \bibnamefont
  {Everts}}\ and\ \bibinfo {author} {\bibfnamefont {B.}~\bibnamefont
  {Cichocki}},\ }\bibfield  {title} {\bibinfo {title} {Dissipative effects in
  odd viscous {Stokes} flow around a single sphere},\ }\href
  {https://doi.org/10.1103/PhysRevLett.132.218303} {\bibfield  {journal}
  {\bibinfo  {journal} {Phys. Rev. Lett.}\ }\textbf {\bibinfo {volume} {132}},\
  \bibinfo {pages} {218303} (\bibinfo {year} {2024})}\BibitemShut {NoStop}%
\bibitem [{\citenamefont {Hosaka}\ \emph {et~al.}(2024)\citenamefont {Hosaka},
  \citenamefont {Chatzittofi}, \citenamefont {Golestanian},\ and\ \citenamefont
  {Vilfan}}]{Hosaka2024}%
  \BibitemOpen
  \bibfield  {author} {\bibinfo {author} {\bibfnamefont {Y.}~\bibnamefont
  {Hosaka}}, \bibinfo {author} {\bibfnamefont {M.}~\bibnamefont {Chatzittofi}},
  \bibinfo {author} {\bibfnamefont {R.}~\bibnamefont {Golestanian}},\ and\
  \bibinfo {author} {\bibfnamefont {A.}~\bibnamefont {Vilfan}},\ }\bibfield
  {title} {\bibinfo {title} {Chirotactic response of microswimmers in fluids
  with odd viscosity},\ }\href
  {https://doi.org/10.1103/PhysRevResearch.6.L032044} {\bibfield  {journal}
  {\bibinfo  {journal} {Phys. Rev. Res.}\ }\textbf {\bibinfo {volume} {6}},\
  \bibinfo {pages} {L032044} (\bibinfo {year} {2024})}\BibitemShut {NoStop}%
\bibitem [{\citenamefont {Read}\ and\ \citenamefont {Rezayi}(2011)}]{Read2011}%
  \BibitemOpen
  \bibfield  {author} {\bibinfo {author} {\bibfnamefont {N.}~\bibnamefont
  {Read}}\ and\ \bibinfo {author} {\bibfnamefont {E.~H.}\ \bibnamefont
  {Rezayi}},\ }\bibfield  {title} {\bibinfo {title} {Hall viscosity, orbital
  spin, and geometry: Paired superfluids and quantum {Hall} systems},\ }\href
  {https://doi.org/10.1103/PhysRevB.84.085316} {\bibfield  {journal} {\bibinfo
  {journal} {Phys. Rev. B}\ }\textbf {\bibinfo {volume} {84}},\ \bibinfo
  {pages} {085316} (\bibinfo {year} {2011})}\BibitemShut {NoStop}%
\bibitem [{\citenamefont {Haldane}()}]{Haldane2009}%
  \BibitemOpen
  \bibfield  {author} {\bibinfo {author} {\bibfnamefont {F.}~\bibnamefont
  {Haldane}},\ }\href@noop {} {\bibinfo {title} {{"Hall viscosity"} and
  intrinsic metric of incompressible fractional {Hall} fluids}},\ \Eprint
  {https://arxiv.org/abs/0906.1854} {arXiv:0906.1854} \BibitemShut {NoStop}%
\bibitem [{\citenamefont {Hoyos}\ and\ \citenamefont {Son}(2012)}]{Hoyos2012}%
  \BibitemOpen
  \bibfield  {author} {\bibinfo {author} {\bibfnamefont {C.}~\bibnamefont
  {Hoyos}}\ and\ \bibinfo {author} {\bibfnamefont {D.~T.}\ \bibnamefont
  {Son}},\ }\bibfield  {title} {\bibinfo {title} {Hall viscosity and
  electromagnetic response},\ }\href
  {https://doi.org/10.1103/PhysRevLett.108.066805} {\bibfield  {journal}
  {\bibinfo  {journal} {Phys. Rev. Lett.}\ }\textbf {\bibinfo {volume} {108}},\
  \bibinfo {pages} {066805} (\bibinfo {year} {2012})}\BibitemShut {NoStop}%
\bibitem [{\citenamefont {Bradlyn}\ \emph {et~al.}(2012)\citenamefont
  {Bradlyn}, \citenamefont {Goldstein},\ and\ \citenamefont
  {Read}}]{Bradlyn2012}%
  \BibitemOpen
  \bibfield  {author} {\bibinfo {author} {\bibfnamefont {B.}~\bibnamefont
  {Bradlyn}}, \bibinfo {author} {\bibfnamefont {M.}~\bibnamefont {Goldstein}},\
  and\ \bibinfo {author} {\bibfnamefont {N.}~\bibnamefont {Read}},\ }\bibfield
  {title} {\bibinfo {title} {Kubo formulas for viscosity: {Hall} viscosity,
  {Ward} identities, and the relation with conductivity},\ }\href
  {https://doi.org/10.1103/PhysRevB.86.245309} {\bibfield  {journal} {\bibinfo
  {journal} {Phys. Rev. B}\ }\textbf {\bibinfo {volume} {86}},\ \bibinfo
  {pages} {245309} (\bibinfo {year} {2012})}\BibitemShut {NoStop}%
\bibitem [{\citenamefont {Alekseev}(2016)}]{Alekseev2016}%
  \BibitemOpen
  \bibfield  {author} {\bibinfo {author} {\bibfnamefont {P.~S.}\ \bibnamefont
  {Alekseev}},\ }\bibfield  {title} {\bibinfo {title} {Negative
  magnetoresistance in viscous flow of two-dimensional electrons},\ }\href
  {https://doi.org/10.1103/PhysRevLett.117.166601} {\bibfield  {journal}
  {\bibinfo  {journal} {Phys. Rev. Lett.}\ }\textbf {\bibinfo {volume} {117}},\
  \bibinfo {pages} {166601} (\bibinfo {year} {2016})}\BibitemShut {NoStop}%
\bibitem [{\citenamefont {Scaffidi}\ \emph {et~al.}(2017)\citenamefont
  {Scaffidi}, \citenamefont {Nandi}, \citenamefont {Schmidt}, \citenamefont
  {Mackenzie},\ and\ \citenamefont {Moore}}]{Scaffidi2017}%
  \BibitemOpen
  \bibfield  {author} {\bibinfo {author} {\bibfnamefont {T.}~\bibnamefont
  {Scaffidi}}, \bibinfo {author} {\bibfnamefont {N.}~\bibnamefont {Nandi}},
  \bibinfo {author} {\bibfnamefont {B.}~\bibnamefont {Schmidt}}, \bibinfo
  {author} {\bibfnamefont {A.~P.}\ \bibnamefont {Mackenzie}},\ and\ \bibinfo
  {author} {\bibfnamefont {J.~E.}\ \bibnamefont {Moore}},\ }\bibfield  {title}
  {\bibinfo {title} {Hydrodynamic electron flow and {Hall} viscosity},\ }\href
  {https://doi.org/10.1103/PhysRevLett.118.226601} {\bibfield  {journal}
  {\bibinfo  {journal} {Phys. Rev. Lett.}\ }\textbf {\bibinfo {volume} {118}},\
  \bibinfo {pages} {226601} (\bibinfo {year} {2017})}\BibitemShut {NoStop}%
\bibitem [{\citenamefont {Delacr{\'e}taz}\ and\ \citenamefont
  {Gromov}(2017)}]{Delacretaz2017}%
  \BibitemOpen
  \bibfield  {author} {\bibinfo {author} {\bibfnamefont {L.~V.}\ \bibnamefont
  {Delacr{\'e}taz}}\ and\ \bibinfo {author} {\bibfnamefont {A.}~\bibnamefont
  {Gromov}},\ }\bibfield  {title} {\bibinfo {title} {Transport signatures of
  the {Hall} viscosity},\ }\href
  {https://doi.org/10.1103/PhysRevLett.119.226602} {\bibfield  {journal}
  {\bibinfo  {journal} {Phys. Rev. Lett.}\ }\textbf {\bibinfo {volume} {119}},\
  \bibinfo {pages} {226602} (\bibinfo {year} {2017})}\BibitemShut {NoStop}%
\bibitem [{\citenamefont {Holder}\ \emph {et~al.}(2019)\citenamefont {Holder},
  \citenamefont {Queiroz},\ and\ \citenamefont {Stern}}]{Holder2019}%
  \BibitemOpen
  \bibfield  {author} {\bibinfo {author} {\bibfnamefont {T.}~\bibnamefont
  {Holder}}, \bibinfo {author} {\bibfnamefont {R.}~\bibnamefont {Queiroz}},\
  and\ \bibinfo {author} {\bibfnamefont {A.}~\bibnamefont {Stern}},\ }\bibfield
   {title} {\bibinfo {title} {Unified description of the classical {Hall}
  viscosity},\ }\href {https://doi.org/10.1103/PhysRevLett.123.106801}
  {\bibfield  {journal} {\bibinfo  {journal} {Phys. Rev. Lett.}\ }\textbf
  {\bibinfo {volume} {123}},\ \bibinfo {pages} {106801} (\bibinfo {year}
  {2019})}\BibitemShut {NoStop}%
\bibitem [{\citenamefont {Afanasiev}\ \emph {et~al.}(2022)\citenamefont
  {Afanasiev}, \citenamefont {Alekseev}, \citenamefont {Danilenko},
  \citenamefont {Dmitriev}, \citenamefont {Greshnov},\ and\ \citenamefont
  {Semina}}]{Afanasiev2022}%
  \BibitemOpen
  \bibfield  {author} {\bibinfo {author} {\bibfnamefont {A.~N.}\ \bibnamefont
  {Afanasiev}}, \bibinfo {author} {\bibfnamefont {P.~S.}\ \bibnamefont
  {Alekseev}}, \bibinfo {author} {\bibfnamefont {A.~A.}\ \bibnamefont
  {Danilenko}}, \bibinfo {author} {\bibfnamefont {A.~P.}\ \bibnamefont
  {Dmitriev}}, \bibinfo {author} {\bibfnamefont {A.~A.}\ \bibnamefont
  {Greshnov}},\ and\ \bibinfo {author} {\bibfnamefont {M.~A.}\ \bibnamefont
  {Semina}},\ }\bibfield  {title} {\bibinfo {title} {Hall effect in
  {Poiseuille} flow of two-dimensional electron fluid},\ }\href
  {https://doi.org/10.1103/PhysRevB.106.245415} {\bibfield  {journal} {\bibinfo
   {journal} {Phys. Rev. B}\ }\textbf {\bibinfo {volume} {106}},\ \bibinfo
  {pages} {245415} (\bibinfo {year} {2022})}\BibitemShut {NoStop}%
\bibitem [{\citenamefont {Ganeshan}\ and\ \citenamefont
  {Abanov}(2017)}]{Ganeshan2017}%
  \BibitemOpen
  \bibfield  {author} {\bibinfo {author} {\bibfnamefont {S.}~\bibnamefont
  {Ganeshan}}\ and\ \bibinfo {author} {\bibfnamefont {A.~G.}\ \bibnamefont
  {Abanov}},\ }\bibfield  {title} {\bibinfo {title} {Odd viscosity in
  two-dimensional incompressible fluids},\ }\href
  {https://doi.org/10.1103/PhysRevFluids.2.094101} {\bibfield  {journal}
  {\bibinfo  {journal} {Phys. Rev. Fluids}\ }\textbf {\bibinfo {volume} {2}},\
  \bibinfo {pages} {094101} (\bibinfo {year} {2017})}\BibitemShut {NoStop}%
\bibitem [{\citenamefont {Alekseev}\ and\ \citenamefont
  {Dmitriev}(2023)}]{Alekseev2023}%
  \BibitemOpen
  \bibfield  {author} {\bibinfo {author} {\bibfnamefont {P.~S.}\ \bibnamefont
  {Alekseev}}\ and\ \bibinfo {author} {\bibfnamefont {A.~P.}\ \bibnamefont
  {Dmitriev}},\ }\bibfield  {title} {\bibinfo {title} {Hydrodynamic
  magnetotransport in two-dimensional electron systems with macroscopic
  obstacles},\ }\href {https://doi.org/10.1103/PhysRevB.108.205413} {\bibfield
  {journal} {\bibinfo  {journal} {Phys. Rev. B}\ }\textbf {\bibinfo {volume}
  {108}},\ \bibinfo {pages} {205413} (\bibinfo {year} {2023})}\BibitemShut
  {NoStop}%
\bibitem [{\citenamefont {Gornyi}\ and\ \citenamefont
  {Polyakov}(2023)}]{Gornyi2023}%
  \BibitemOpen
  \bibfield  {author} {\bibinfo {author} {\bibfnamefont {I.~V.}\ \bibnamefont
  {Gornyi}}\ and\ \bibinfo {author} {\bibfnamefont {D.~G.}\ \bibnamefont
  {Polyakov}},\ }\bibfield  {title} {\bibinfo {title} {Two-dimensional electron
  hydrodynamics in a random array of impenetrable obstacles:
  {Magnetoresistivity}, {Hall} viscosity, and the {Landauer} dipole},\ }\href
  {https://doi.org/10.1103/PhysRevB.108.165429} {\bibfield  {journal} {\bibinfo
   {journal} {Phys. Rev. B}\ }\textbf {\bibinfo {volume} {108}},\ \bibinfo
  {pages} {165429} (\bibinfo {year} {2023})}\BibitemShut {NoStop}%
\bibitem [{\citenamefont {Messica}\ \emph {et~al.}(2023)\citenamefont
  {Messica}, \citenamefont {Gutman},\ and\ \citenamefont
  {Ostrovsky}}]{Messica2023}%
  \BibitemOpen
  \bibfield  {author} {\bibinfo {author} {\bibfnamefont {Y.}~\bibnamefont
  {Messica}}, \bibinfo {author} {\bibfnamefont {D.~B.}\ \bibnamefont
  {Gutman}},\ and\ \bibinfo {author} {\bibfnamefont {P.~M.}\ \bibnamefont
  {Ostrovsky}},\ }\bibfield  {title} {\bibinfo {title} {Anomalous {Hall} effect
  in disordered {Weyl} semimetals},\ }\href
  {https://doi.org/10.1103/PhysRevB.108.045121} {\bibfield  {journal} {\bibinfo
   {journal} {Phys. Rev. B}\ }\textbf {\bibinfo {volume} {108}},\ \bibinfo
  {pages} {045121} (\bibinfo {year} {2023})}\BibitemShut {NoStop}%
\bibitem [{\citenamefont {Krebs}\ \emph {et~al.}(2023)\citenamefont {Krebs},
  \citenamefont {Behn}, \citenamefont {Li}, \citenamefont {Smith},
  \citenamefont {Watanabe}, \citenamefont {Taniguchi}, \citenamefont
  {Levchenko},\ and\ \citenamefont {Brar}}]{Krebs2023}%
  \BibitemOpen
  \bibfield  {author} {\bibinfo {author} {\bibfnamefont {Z.~J.}\ \bibnamefont
  {Krebs}}, \bibinfo {author} {\bibfnamefont {W.~A.}\ \bibnamefont {Behn}},
  \bibinfo {author} {\bibfnamefont {S.}~\bibnamefont {Li}}, \bibinfo {author}
  {\bibfnamefont {K.~J.}\ \bibnamefont {Smith}}, \bibinfo {author}
  {\bibfnamefont {K.}~\bibnamefont {Watanabe}}, \bibinfo {author}
  {\bibfnamefont {T.}~\bibnamefont {Taniguchi}}, \bibinfo {author}
  {\bibfnamefont {A.}~\bibnamefont {Levchenko}},\ and\ \bibinfo {author}
  {\bibfnamefont {V.~W.}\ \bibnamefont {Brar}},\ }\bibfield  {title} {\bibinfo
  {title} {Imaging the breaking of electrostatic dams in graphene for ballistic
  and viscous fluids},\ }\href {https://doi.org/10.1126/science.abm6073}
  {\bibfield  {journal} {\bibinfo  {journal} {Science}\ }\textbf {\bibinfo
  {volume} {379}},\ \bibinfo {pages} {671} (\bibinfo {year}
  {2023})}\BibitemShut {NoStop}%
\bibitem [{\citenamefont {Ikegami}\ \emph {et~al.}(2013)\citenamefont
  {Ikegami}, \citenamefont {Tsutsumi},\ and\ \citenamefont
  {Kono}}]{Ikegami2013}%
  \BibitemOpen
  \bibfield  {author} {\bibinfo {author} {\bibfnamefont {H.}~\bibnamefont
  {Ikegami}}, \bibinfo {author} {\bibfnamefont {Y.}~\bibnamefont {Tsutsumi}},\
  and\ \bibinfo {author} {\bibfnamefont {K.}~\bibnamefont {Kono}},\ }\bibfield
  {title} {\bibinfo {title} {Chiral symmetry breaking in superfluid
  $^3${He-A}},\ }\href {https://doi.org/10.1126/science.1236509} {\bibfield
  {journal} {\bibinfo  {journal} {Science}\ }\textbf {\bibinfo {volume}
  {341}},\ \bibinfo {pages} {59} (\bibinfo {year} {2013})}\BibitemShut
  {NoStop}%
\bibitem [{\citenamefont {Kwon}(2006)}]{Kwon2006}%
  \BibitemOpen
  \bibfield  {author} {\bibinfo {author} {\bibfnamefont {Y.~D.}\ \bibnamefont
  {Kwon}},\ }\bibfield  {title} {\bibinfo {title} {Theory of the screened
  {Coulomb} field generated by impurity ions in semiconductors},\ }\href
  {https://doi.org/10.1103/PhysRevB.73.165210} {\bibfield  {journal} {\bibinfo
  {journal} {Phys. Rev. B}\ }\textbf {\bibinfo {volume} {73}},\ \bibinfo
  {pages} {165210} (\bibinfo {year} {2006})}\BibitemShut {NoStop}%
\bibitem [{\citenamefont {Son}\ and\ \citenamefont {Spivak}(2013)}]{Son2013}%
  \BibitemOpen
  \bibfield  {author} {\bibinfo {author} {\bibfnamefont {D.}~\bibnamefont
  {Son}}\ and\ \bibinfo {author} {\bibfnamefont {B.}~\bibnamefont {Spivak}},\
  }\bibfield  {title} {\bibinfo {title} {Chiral anomaly and classical negative
  magnetoresistance of {Weyl} metals},\ }\href
  {https://doi.org/10.1103/PhysRevB.88.104412} {\bibfield  {journal} {\bibinfo
  {journal} {Phys. Rev. B}\ }\textbf {\bibinfo {volume} {88}},\ \bibinfo
  {pages} {104412} (\bibinfo {year} {2013})}\BibitemShut {NoStop}%
\bibitem [{\citenamefont {Landsteiner}\ \emph {et~al.}(2015)\citenamefont
  {Landsteiner}, \citenamefont {Liu},\ and\ \citenamefont
  {Sun}}]{Landsteiner2015}%
  \BibitemOpen
  \bibfield  {author} {\bibinfo {author} {\bibfnamefont {K.}~\bibnamefont
  {Landsteiner}}, \bibinfo {author} {\bibfnamefont {Y.}~\bibnamefont {Liu}},\
  and\ \bibinfo {author} {\bibfnamefont {Y.~W.}\ \bibnamefont {Sun}},\
  }\bibfield  {title} {\bibinfo {title} {Negative magnetoresistivity in chiral
  fluids and holography},\ }\href {https://doi.org/10.1007/JHEP03(2015)127}
  {\bibfield  {journal} {\bibinfo  {journal} {J. High Energy Phys.}\ }\textbf
  {\bibinfo {volume} {2015}},\ \bibinfo {pages} {1}}\BibitemShut {NoStop}%
\bibitem [{\citenamefont {Lucas}\ \emph {et~al.}(2016)\citenamefont {Lucas},
  \citenamefont {Davison},\ and\ \citenamefont {Sachdev}}]{Lucas2016}%
  \BibitemOpen
  \bibfield  {author} {\bibinfo {author} {\bibfnamefont {A.}~\bibnamefont
  {Lucas}}, \bibinfo {author} {\bibfnamefont {R.~A.}\ \bibnamefont {Davison}},\
  and\ \bibinfo {author} {\bibfnamefont {S.}~\bibnamefont {Sachdev}},\
  }\bibfield  {title} {\bibinfo {title} {Hydrodynamic theory of thermoelectric
  transport and negative magnetoresistance in {Weyl} semimetals},\ }\href
  {https://doi.org/10.1073/pnas.1608881113} {\bibfield  {journal} {\bibinfo
  {journal} {Proc. Natl. Acad. Sci. U. S. A.}\ }\textbf {\bibinfo {volume}
  {113}},\ \bibinfo {pages} {9463} (\bibinfo {year} {2016})}\BibitemShut
  {NoStop}%
\bibitem [{\citenamefont {Gorbar}\ \emph {et~al.}(2018)\citenamefont {Gorbar},
  \citenamefont {Miransky}, \citenamefont {Shovkovy},\ and\ \citenamefont
  {Sukhachov}}]{Gorbar2018}%
  \BibitemOpen
  \bibfield  {author} {\bibinfo {author} {\bibfnamefont {E.~V.}\ \bibnamefont
  {Gorbar}}, \bibinfo {author} {\bibfnamefont {V.~A.}\ \bibnamefont
  {Miransky}}, \bibinfo {author} {\bibfnamefont {I.~A.}\ \bibnamefont
  {Shovkovy}},\ and\ \bibinfo {author} {\bibfnamefont {P.~O.}\ \bibnamefont
  {Sukhachov}},\ }\bibfield  {title} {\bibinfo {title} {Consistent hydrodynamic
  theory of chiral electrons in {Weyl} semimetals},\ }\href
  {https://journals.aps.org/prb/abstract/10.1103/PhysRevB.97.121105} {\bibfield
   {journal} {\bibinfo  {journal} {Phys. Rev. B}\ }\textbf {\bibinfo {volume}
  {97}},\ \bibinfo {pages} {121105(R)} (\bibinfo {year} {2018})}\BibitemShut
  {NoStop}%
\bibitem [{\citenamefont {Sukhachov}\ \emph {et~al.}(2018)\citenamefont
  {Sukhachov}, \citenamefont {Gorbar}, \citenamefont {Shovkovy},\ and\
  \citenamefont {Miransky}}]{Sukhachov2018}%
  \BibitemOpen
  \bibfield  {author} {\bibinfo {author} {\bibfnamefont {P.~O.}\ \bibnamefont
  {Sukhachov}}, \bibinfo {author} {\bibfnamefont {E.~V.}\ \bibnamefont
  {Gorbar}}, \bibinfo {author} {\bibfnamefont {I.~A.}\ \bibnamefont
  {Shovkovy}},\ and\ \bibinfo {author} {\bibfnamefont {V.~A.}\ \bibnamefont
  {Miransky}},\ }\bibfield  {title} {\bibinfo {title} {Collective excitations
  in {Weyl} semimetals in the hydrodynamic regime},\ }\href
  {https://doi.org/10.1088/1361-648X/aac500} {\bibfield  {journal} {\bibinfo
  {journal} {J. Condens. Matter Phys.}\ }\textbf {\bibinfo {volume} {30}},\
  \bibinfo {pages} {275601} (\bibinfo {year} {2018})}\BibitemShut {NoStop}%
\bibitem [{\citenamefont {Narozhny}(2019)}]{Narozhny2019}%
  \BibitemOpen
  \bibfield  {author} {\bibinfo {author} {\bibfnamefont {B.~N.}\ \bibnamefont
  {Narozhny}},\ }\bibfield  {title} {\bibinfo {title} {Electronic hydrodynamics
  in graphene},\ }\href {https://doi.org/10.1016/J.AOP.2019.167979} {\bibfield
  {journal} {\bibinfo  {journal} {Ann. Phys.}\ }\textbf {\bibinfo {volume}
  {411}},\ \bibinfo {pages} {167979} (\bibinfo {year} {2019})}\BibitemShut
  {NoStop}%
\bibitem [{\citenamefont {Narozhny}(2022)}]{Narozhny2022}%
  \BibitemOpen
  \bibfield  {author} {\bibinfo {author} {\bibfnamefont {B.~N.}\ \bibnamefont
  {Narozhny}},\ }\bibfield  {title} {\bibinfo {title} {Hydrodynamic approach to
  two-dimensional electron systems},\ }\href
  {https://doi.org/10.1007/s40766-022-00036-z} {\bibfield  {journal} {\bibinfo
  {journal} {Riv. Nuovo Cim.}\ }\textbf {\bibinfo {volume} {45}},\ \bibinfo
  {pages} {661} (\bibinfo {year} {2022})}\BibitemShut {NoStop}%
\bibitem [{\citenamefont {Fritz}\ and\ \citenamefont
  {Scaffidi}(2024)}]{Fritz2024}%
  \BibitemOpen
  \bibfield  {author} {\bibinfo {author} {\bibfnamefont {L.}~\bibnamefont
  {Fritz}}\ and\ \bibinfo {author} {\bibfnamefont {T.}~\bibnamefont
  {Scaffidi}},\ }\bibfield  {title} {\bibinfo {title} {Hydrodynamic electronic
  transport},\ }\href
  {https://doi.org/10.1146/annurev-conmatphys-040521-042014} {\bibfield
  {journal} {\bibinfo  {journal} {Annu. Rev. Condens. Matter Phys.}\ }\textbf
  {\bibinfo {volume} {15}},\ \bibinfo {pages} {17} (\bibinfo {year}
  {2024})}\BibitemShut {NoStop}%
\bibitem [{\citenamefont {Landau}\ and\ \citenamefont
  {Lifshitz}(1987)}]{Landau2003}%
  \BibitemOpen
  \bibfield  {author} {\bibinfo {author} {\bibfnamefont {L.~D.}\ \bibnamefont
  {Landau}}\ and\ \bibinfo {author} {\bibfnamefont {E.~M.}\ \bibnamefont
  {Lifshitz}},\ }\href@noop {} {\emph {\bibinfo {title} {Fluid Mechanics}}},\
  \bibinfo {edition} {2nd}\ ed.,\ \bibinfo {series} {Course of Theoretical
  Physics}, Vol.~\bibinfo {volume} {6}\ (\bibinfo  {publisher} {Elsevier},\
  \bibinfo {address} {New York},\ \bibinfo {year} {1987})\BibitemShut {NoStop}%
\bibitem [{\citenamefont {Haldane}(2004)}]{Haldane2004}%
  \BibitemOpen
  \bibfield  {author} {\bibinfo {author} {\bibfnamefont {F.~D.}\ \bibnamefont
  {Haldane}},\ }\bibfield  {title} {\bibinfo {title} {Berry curvature on the
  {Fermi} surface: Anomalous {Hall} effect as a topological {Fermi}-liquid
  property},\ }\href {https://doi.org/10.1103/PhysRevLett.93.206602} {\bibfield
   {journal} {\bibinfo  {journal} {Phys. Rev. Lett.}\ }\textbf {\bibinfo
  {volume} {93}},\ \bibinfo {pages} {206602} (\bibinfo {year}
  {2004})}\BibitemShut {NoStop}%
\bibitem [{\citenamefont {Nagaosa}\ \emph {et~al.}(2010)\citenamefont
  {Nagaosa}, \citenamefont {Sinova}, \citenamefont {Onoda}, \citenamefont
  {MacDonald},\ and\ \citenamefont {Ong}}]{Nagaosa2010}%
  \BibitemOpen
  \bibfield  {author} {\bibinfo {author} {\bibfnamefont {N.}~\bibnamefont
  {Nagaosa}}, \bibinfo {author} {\bibfnamefont {J.}~\bibnamefont {Sinova}},
  \bibinfo {author} {\bibfnamefont {S.}~\bibnamefont {Onoda}}, \bibinfo
  {author} {\bibfnamefont {A.~H.}\ \bibnamefont {MacDonald}},\ and\ \bibinfo
  {author} {\bibfnamefont {N.~P.}\ \bibnamefont {Ong}},\ }\bibfield  {title}
  {\bibinfo {title} {Anomalous {Hall} effect},\ }\href
  {https://doi.org/10.1103/RevModPhys.82.1539} {\bibfield  {journal} {\bibinfo
  {journal} {RMP}\ }\textbf {\bibinfo {volume} {82}},\ \bibinfo {pages} {1539}
  (\bibinfo {year} {2010})}\BibitemShut {NoStop}%
\bibitem [{\citenamefont {Yan}\ and\ \citenamefont {Felser}(2017)}]{Yan2017}%
  \BibitemOpen
  \bibfield  {author} {\bibinfo {author} {\bibfnamefont {B.}~\bibnamefont
  {Yan}}\ and\ \bibinfo {author} {\bibfnamefont {C.}~\bibnamefont {Felser}},\
  }\bibfield  {title} {\bibinfo {title} {Topological materials: {Weyl}
  semimetals},\ }\href
  {https://doi.org/10.1146/annurev-conmatphys-031016-025458} {\bibfield
  {journal} {\bibinfo  {journal} {Annu. Rev. Condens. Matter Phys.}\ }\textbf
  {\bibinfo {volume} {8}},\ \bibinfo {pages} {337} (\bibinfo {year}
  {2017})}\BibitemShut {NoStop}%
\bibitem [{\citenamefont {Armitage}\ \emph {et~al.}(2018)\citenamefont
  {Armitage}, \citenamefont {Mele},\ and\ \citenamefont
  {Vishwanath}}]{Armitage2018}%
  \BibitemOpen
  \bibfield  {author} {\bibinfo {author} {\bibfnamefont {N.~P.}\ \bibnamefont
  {Armitage}}, \bibinfo {author} {\bibfnamefont {E.~J.}\ \bibnamefont {Mele}},\
  and\ \bibinfo {author} {\bibfnamefont {A.}~\bibnamefont {Vishwanath}},\
  }\bibfield  {title} {\bibinfo {title} {Weyl and {Dirac} semimetals in
  three-dimensional solids},\ }\href
  {https://doi.org/10.1103/RevModPhys.90.015001} {\bibfield  {journal}
  {\bibinfo  {journal} {RMP}\ }\textbf {\bibinfo {volume} {90}},\ \bibinfo
  {pages} {015001} (\bibinfo {year} {2018})}\BibitemShut {NoStop}%
\bibitem [{\citenamefont {Gooth}\ \emph {et~al.}(2018)\citenamefont {Gooth},
  \citenamefont {Menges}, \citenamefont {Kumar}, \citenamefont {S{\"u}{\ss}},
  \citenamefont {Shekhar}, \citenamefont {Sun}, \citenamefont {Drechsler},
  \citenamefont {Zierold}, \citenamefont {Felser},\ and\ \citenamefont
  {Gotsmann}}]{Gooth2018}%
  \BibitemOpen
  \bibfield  {author} {\bibinfo {author} {\bibfnamefont {J.}~\bibnamefont
  {Gooth}}, \bibinfo {author} {\bibfnamefont {F.}~\bibnamefont {Menges}},
  \bibinfo {author} {\bibfnamefont {N.}~\bibnamefont {Kumar}}, \bibinfo
  {author} {\bibfnamefont {V.}~\bibnamefont {S{\"u}{\ss}}}, \bibinfo {author}
  {\bibfnamefont {C.}~\bibnamefont {Shekhar}}, \bibinfo {author} {\bibfnamefont
  {Y.}~\bibnamefont {Sun}}, \bibinfo {author} {\bibfnamefont {U.}~\bibnamefont
  {Drechsler}}, \bibinfo {author} {\bibfnamefont {R.}~\bibnamefont {Zierold}},
  \bibinfo {author} {\bibfnamefont {C.}~\bibnamefont {Felser}},\ and\ \bibinfo
  {author} {\bibfnamefont {B.}~\bibnamefont {Gotsmann}},\ }\bibfield  {title}
  {\bibinfo {title} {Thermal and electrical signatures of a hydrodynamic
  electron fluid in tungsten diphosphide},\ }\href
  {https://doi.org/10.1038/s41467-018-06688-y} {\bibfield  {journal} {\bibinfo
  {journal} {Nat. Commun.}\ }\textbf {\bibinfo {volume} {9}},\ \bibinfo {pages}
  {4093} (\bibinfo {year} {2018})}\BibitemShut {NoStop}%
\bibitem [{\citenamefont {Jaoui}\ \emph {et~al.}(2018)\citenamefont {Jaoui},
  \citenamefont {Fauqu{\'e}}, \citenamefont {Rischau}, \citenamefont {Subedi},
  \citenamefont {Fu}, \citenamefont {Gooth}, \citenamefont {Kumar},
  \citenamefont {S{\"u}{\ss}}, \citenamefont {Maslov}, \citenamefont {Felser},\
  and\ \citenamefont {Kamran}}]{Jaoui2018}%
  \BibitemOpen
  \bibfield  {author} {\bibinfo {author} {\bibfnamefont {A.}~\bibnamefont
  {Jaoui}}, \bibinfo {author} {\bibfnamefont {B.}~\bibnamefont {Fauqu{\'e}}},
  \bibinfo {author} {\bibfnamefont {C.~W.}\ \bibnamefont {Rischau}}, \bibinfo
  {author} {\bibfnamefont {A.}~\bibnamefont {Subedi}}, \bibinfo {author}
  {\bibfnamefont {C.}~\bibnamefont {Fu}}, \bibinfo {author} {\bibfnamefont
  {J.}~\bibnamefont {Gooth}}, \bibinfo {author} {\bibfnamefont
  {N.}~\bibnamefont {Kumar}}, \bibinfo {author} {\bibfnamefont
  {V.}~\bibnamefont {S{\"u}{\ss}}}, \bibinfo {author} {\bibfnamefont {D.~L.}\
  \bibnamefont {Maslov}}, \bibinfo {author} {\bibfnamefont {C.}~\bibnamefont
  {Felser}},\ and\ \bibinfo {author} {\bibfnamefont {B.}~\bibnamefont
  {Kamran}},\ }\bibfield  {title} {\bibinfo {title} {Departure from the
  {Wiedemann-Franz} law in {WP}$_2$ driven by mismatch in {$T$}-square
  resistivity prefactors},\ }\href {https://doi.org/10.1038/s41535-018-0136-x}
  {\bibfield  {journal} {\bibinfo  {journal} {npj Quantum Mater.}\ }\textbf
  {\bibinfo {volume} {3}},\ \bibinfo {pages} {64} (\bibinfo {year}
  {2018})}\BibitemShut {NoStop}%
\bibitem [{\citenamefont {Burkov}\ and\ \citenamefont
  {Balents}(2011)}]{Burkov2011}%
  \BibitemOpen
  \bibfield  {author} {\bibinfo {author} {\bibfnamefont {A.~A.}\ \bibnamefont
  {Burkov}}\ and\ \bibinfo {author} {\bibfnamefont {L.}~\bibnamefont
  {Balents}},\ }\bibfield  {title} {\bibinfo {title} {Weyl semimetal in a
  topological insulator multilayer},\ }\href
  {https://doi.org/10.1103/PhysRevLett.107.127205} {\bibfield  {journal}
  {\bibinfo  {journal} {Phys. Rev. Lett.}\ }\textbf {\bibinfo {volume} {107}},\
  \bibinfo {pages} {127205} (\bibinfo {year} {2011})}\BibitemShut {NoStop}%
\bibitem [{\citenamefont {Lier}(2024)}]{Lier2024}%
  \BibitemOpen
  \bibfield  {author} {\bibinfo {author} {\bibfnamefont {R.}~\bibnamefont
  {Lier}},\ }\bibfield  {title} {\bibinfo {title} {Odd viscous flow past a
  sphere at low but non-zero {Reynolds} numbers},\ }\href
  {https://doi.org/10.1017/jfm.2024.915} {\bibfield  {journal} {\bibinfo
  {journal} {J. Fluid Mech.}\ }\textbf {\bibinfo {volume} {998}},\ \bibinfo
  {pages} {A40} (\bibinfo {year} {2024})}\BibitemShut {NoStop}%
\bibitem [{\citenamefont {Stokes}(1851)}]{Stokes1851}%
  \BibitemOpen
  \bibfield  {author} {\bibinfo {author} {\bibfnamefont {G.~G.}\ \bibnamefont
  {Stokes}},\ }\bibfield  {title} {\bibinfo {title} {On the effect of the
  internal friction of fluids on the motion of pendulums},\ }\href
  {https://doi.org/10.1017/cbo9780511702266.002} {\bibfield  {journal}
  {\bibinfo  {journal} {Trans. Cambridge Philos. Soc.}\ }\textbf {\bibinfo
  {volume} {9}},\ \bibinfo {pages} {8} (\bibinfo {year} {1851})}\BibitemShut
  {NoStop}%
\bibitem [{\citenamefont {Levitov}\ and\ \citenamefont
  {Falkovich}(2016)}]{Levitov2016}%
  \BibitemOpen
  \bibfield  {author} {\bibinfo {author} {\bibfnamefont {L.}~\bibnamefont
  {Levitov}}\ and\ \bibinfo {author} {\bibfnamefont {G.}~\bibnamefont
  {Falkovich}},\ }\bibfield  {title} {\bibinfo {title} {Electron viscosity,
  current vortices and negative nonlocal resistance in graphene},\ }\href
  {https://doi.org/10.1038/nphys3667} {\bibfield  {journal} {\bibinfo
  {journal} {Nat. Phys.}\ }\textbf {\bibinfo {volume} {12}},\ \bibinfo {pages}
  {672} (\bibinfo {year} {2016})}\BibitemShut {NoStop}%
\bibitem [{\citenamefont {Yang}\ \emph {et~al.}(2011)\citenamefont {Yang},
  \citenamefont {Lu},\ and\ \citenamefont {Ran}}]{Yang2011}%
  \BibitemOpen
  \bibfield  {author} {\bibinfo {author} {\bibfnamefont {K.~Y.}\ \bibnamefont
  {Yang}}, \bibinfo {author} {\bibfnamefont {Y.~M.}\ \bibnamefont {Lu}},\ and\
  \bibinfo {author} {\bibfnamefont {Y.}~\bibnamefont {Ran}},\ }\bibfield
  {title} {\bibinfo {title} {Quantum {Hall} effects in a {Weyl} semimetal:
  Possible application in pyrochlore iridates},\ }\href
  {https://doi.org/10.1103/PhysRevB.84.075129} {\bibfield  {journal} {\bibinfo
  {journal} {Phys. Rev. B}\ }\textbf {\bibinfo {volume} {84}},\ \bibinfo
  {pages} {075129} (\bibinfo {year} {2011})}\BibitemShut {NoStop}%
\bibitem [{\citenamefont {Burkov}(2014)}]{Burkov2014}%
  \BibitemOpen
  \bibfield  {author} {\bibinfo {author} {\bibfnamefont {A.~A.}\ \bibnamefont
  {Burkov}},\ }\bibfield  {title} {\bibinfo {title} {Anomalous {Hall} effect in
  {Weyl} metals},\ }\href {https://doi.org/10.1103/PhysRevLett.113.187202}
  {\bibfield  {journal} {\bibinfo  {journal} {Phys. Rev. Lett.}\ }\textbf
  {\bibinfo {volume} {113}},\ \bibinfo {pages} {187202} (\bibinfo {year}
  {2014})}\BibitemShut {NoStop}%
\bibitem [{\citenamefont {Hruska}\ and\ \citenamefont
  {Spivak}(2002)}]{Hruska2002}%
  \BibitemOpen
  \bibfield  {author} {\bibinfo {author} {\bibfnamefont {M.}~\bibnamefont
  {Hruska}}\ and\ \bibinfo {author} {\bibfnamefont {B.}~\bibnamefont
  {Spivak}},\ }\bibfield  {title} {\bibinfo {title} {Conductivity of the
  classical two-dimensional electron gas},\ }\href
  {https://doi.org/10.1103/PhysRevB.65.033315} {\bibfield  {journal} {\bibinfo
  {journal} {Phys. Rev. B}\ }\textbf {\bibinfo {volume} {65}},\ \bibinfo
  {pages} {033315} (\bibinfo {year} {2002})}\BibitemShut {NoStop}%
\bibitem [{\citenamefont {Guo}\ \emph {et~al.}()\citenamefont {Guo},
  \citenamefont {Ilseven}, \citenamefont {Falkovich},\ and\ \citenamefont
  {Levitov}}]{Guo2016}%
  \BibitemOpen
  \bibfield  {author} {\bibinfo {author} {\bibfnamefont {H.}~\bibnamefont
  {Guo}}, \bibinfo {author} {\bibfnamefont {E.}~\bibnamefont {Ilseven}},
  \bibinfo {author} {\bibfnamefont {G.}~\bibnamefont {Falkovich}},\ and\
  \bibinfo {author} {\bibfnamefont {L.}~\bibnamefont {Levitov}},\ }\href@noop
  {} {\bibinfo {title} {Stokes paradox, back reflections and
  interaction-enhanced conduction}},\ \Eprint
  {https://arxiv.org/abs/1612.09239} {arXiv:1612.09239} \BibitemShut {NoStop}%
\bibitem [{\citenamefont {Sinitsyn}\ \emph {et~al.}(2006)\citenamefont
  {Sinitsyn}, \citenamefont {Niu},\ and\ \citenamefont
  {MacDonald}}]{Sinitsyn2006}%
  \BibitemOpen
  \bibfield  {author} {\bibinfo {author} {\bibfnamefont {N.~A.}\ \bibnamefont
  {Sinitsyn}}, \bibinfo {author} {\bibfnamefont {Q.}~\bibnamefont {Niu}},\ and\
  \bibinfo {author} {\bibfnamefont {A.~H.}\ \bibnamefont {MacDonald}},\
  }\bibfield  {title} {\bibinfo {title} {Coordinate shift in the semiclassical
  {Boltzmann} equation and the anomalous {Hall} effect},\ }\href
  {https://doi.org/10.1103/PhysRevB.73.075318} {\bibfield  {journal} {\bibinfo
  {journal} {Phys. Rev. B}\ }\textbf {\bibinfo {volume} {73}},\ \bibinfo
  {pages} {075318} (\bibinfo {year} {2006})}\BibitemShut {NoStop}%
\bibitem [{\citenamefont {Atencia}\ \emph {et~al.}(2022)\citenamefont
  {Atencia}, \citenamefont {Niu},\ and\ \citenamefont {Culcer}}]{Atencia2022}%
  \BibitemOpen
  \bibfield  {author} {\bibinfo {author} {\bibfnamefont {R.~B.}\ \bibnamefont
  {Atencia}}, \bibinfo {author} {\bibfnamefont {Q.}~\bibnamefont {Niu}},\ and\
  \bibinfo {author} {\bibfnamefont {D.}~\bibnamefont {Culcer}},\ }\bibfield
  {title} {\bibinfo {title} {Semiclassical response of disordered conductors:
  Extrinsic carrier velocity and spin and field-corrected collision integral},\
  }\href {https://doi.org/10.1103/PhysRevResearch.4.013001} {\bibfield
  {journal} {\bibinfo  {journal} {Phys. Rev. Research}\ }\textbf {\bibinfo
  {volume} {4}},\ \bibinfo {pages} {013001} (\bibinfo {year}
  {2022})}\BibitemShut {NoStop}%
\bibitem [{\citenamefont {M{\"u}ller}\ \emph {et~al.}(2009)\citenamefont
  {M{\"u}ller}, \citenamefont {Schmalian},\ and\ \citenamefont
  {Fritz}}]{Muller2009}%
  \BibitemOpen
  \bibfield  {author} {\bibinfo {author} {\bibfnamefont {M.}~\bibnamefont
  {M{\"u}ller}}, \bibinfo {author} {\bibfnamefont {J.}~\bibnamefont
  {Schmalian}},\ and\ \bibinfo {author} {\bibfnamefont {L.}~\bibnamefont
  {Fritz}},\ }\bibfield  {title} {\bibinfo {title} {Graphene: A nearly perfect
  fluid},\ }\href {https://doi.org/10.1103/PhysRevLett.103.025301} {\bibfield
  {journal} {\bibinfo  {journal} {Phys. Rev. Lett.}\ }\textbf {\bibinfo
  {volume} {103}},\ \bibinfo {pages} {025301} (\bibinfo {year}
  {2009})}\BibitemShut {NoStop}%
\bibitem [{\citenamefont {Yudson}\ and\ \citenamefont
  {Maslov}(2007)}]{Yudson2007}%
  \BibitemOpen
  \bibfield  {author} {\bibinfo {author} {\bibfnamefont {V.~I.}\ \bibnamefont
  {Yudson}}\ and\ \bibinfo {author} {\bibfnamefont {D.~L.}\ \bibnamefont
  {Maslov}},\ }\bibfield  {title} {\bibinfo {title} {Universality in scattering
  by large-scale potential fluctuations in two-dimensional conductors},\ }\href
  {https://doi.org/10.1103/PhysRevB.75.241408} {\bibfield  {journal} {\bibinfo
  {journal} {Phys. Rev. B}\ }\textbf {\bibinfo {volume} {75}},\ \bibinfo
  {pages} {241408} (\bibinfo {year} {2007})}\BibitemShut {NoStop}%
\bibitem [{\citenamefont {Gantmakher}\ and\ \citenamefont
  {Levinson}(1987)}]{gantmakher1987carrier}%
  \BibitemOpen
  \bibfield  {author} {\bibinfo {author} {\bibfnamefont {V.}~\bibnamefont
  {Gantmakher}}\ and\ \bibinfo {author} {\bibfnamefont {Y.}~\bibnamefont
  {Levinson}},\ }\href@noop {} {\emph {\bibinfo {title} {Carrier scattering in
  metals and semiconductors}}},\ Vol.~\bibinfo {volume} {19}\ (\bibinfo
  {publisher} {North-Holland},\ \bibinfo {address} {Amsterdam},\ \bibinfo
  {year} {1987})\BibitemShut {NoStop}%
\bibitem [{\citenamefont {Principi}\ and\ \citenamefont
  {Vignale}(2015)}]{Principi2015}%
  \BibitemOpen
  \bibfield  {author} {\bibinfo {author} {\bibfnamefont {A.}~\bibnamefont
  {Principi}}\ and\ \bibinfo {author} {\bibfnamefont {G.}~\bibnamefont
  {Vignale}},\ }\bibfield  {title} {\bibinfo {title} {Violation of the
  {Wiedemann-Franz} law in hydrodynamic electron liquids},\ }\href
  {https://doi.org/10.1103/PhysRevLett.115.056603} {\bibfield  {journal}
  {\bibinfo  {journal} {Phys. Rev. Lett.}\ }\textbf {\bibinfo {volume} {115}},\
  \bibinfo {pages} {56603} (\bibinfo {year} {2015})}\BibitemShut {NoStop}%
\bibitem [{\citenamefont {Kiselev}\ and\ \citenamefont
  {Schmalian}(2019)}]{Kiselev2019}%
  \BibitemOpen
  \bibfield  {author} {\bibinfo {author} {\bibfnamefont {E.~I.}\ \bibnamefont
  {Kiselev}}\ and\ \bibinfo {author} {\bibfnamefont {J.}~\bibnamefont
  {Schmalian}},\ }\bibfield  {title} {\bibinfo {title} {Boundary conditions of
  viscous electron flow},\ }\href {https://doi.org/10.1103/PhysRevB.99.035430}
  {\bibfield  {journal} {\bibinfo  {journal} {Phys. Rev. B}\ }\textbf {\bibinfo
  {volume} {99}},\ \bibinfo {pages} {035430} (\bibinfo {year}
  {2019})}\BibitemShut {NoStop}%
\bibitem [{\citenamefont {Moessner}\ \emph {et~al.}(2019)\citenamefont
  {Moessner}, \citenamefont {Morales-Dur{\'a}n}, \citenamefont {Sur{\'o}wka},\
  and\ \citenamefont {Witkowski}}]{Moessner2019}%
  \BibitemOpen
  \bibfield  {author} {\bibinfo {author} {\bibfnamefont {R.}~\bibnamefont
  {Moessner}}, \bibinfo {author} {\bibfnamefont {N.}~\bibnamefont
  {Morales-Dur{\'a}n}}, \bibinfo {author} {\bibfnamefont {P.}~\bibnamefont
  {Sur{\'o}wka}},\ and\ \bibinfo {author} {\bibfnamefont {P.}~\bibnamefont
  {Witkowski}},\ }\bibfield  {title} {\bibinfo {title} {Boundary-condition and
  geometry engineering in electronic hydrodynamics},\ }\href
  {https://doi.org/10.1103/PhysRevB.100.155115} {\bibfield  {journal} {\bibinfo
   {journal} {Phys. Rev. B}\ }\textbf {\bibinfo {volume} {100}},\ \bibinfo
  {pages} {155115} (\bibinfo {year} {2019})}\BibitemShut {NoStop}%
\bibitem [{\citenamefont {Pitaevskii}\ and\ \citenamefont
  {Lifshitz}(2012)}]{LL-V10}%
  \BibitemOpen
  \bibfield  {author} {\bibinfo {author} {\bibfnamefont {L.~P.}\ \bibnamefont
  {Pitaevskii}}\ and\ \bibinfo {author} {\bibfnamefont {E.}~\bibnamefont
  {Lifshitz}},\ }\href@noop {} {\emph {\bibinfo {title} {Physical Kinetics}}},\
  Vol.~\bibinfo {volume} {10}\ (\bibinfo  {publisher} {Elsevier Science},\
  \bibinfo {address} {New York},\ \bibinfo {year} {2012})\BibitemShut {NoStop}%
\bibitem [{\citenamefont {Gurzhi}(1963)}]{Gurzhi1963}%
  \BibitemOpen
  \bibfield  {author} {\bibinfo {author} {\bibfnamefont {R.~N.}\ \bibnamefont
  {Gurzhi}},\ }\bibfield  {title} {\bibinfo {title} {Minimum of resistance in
  impurity-free conductors},\ }\href@noop {} {\bibfield  {journal} {\bibinfo
  {journal} {J. Exp. Theor. Phys.}\ }\textbf {\bibinfo {volume} {44}} (\bibinfo
  {year} {1963})}\BibitemShut {NoStop}%
\bibitem [{\citenamefont {Messica}\ and\ \citenamefont
  {Gutman}(2024)}]{Messica2024}%
  \BibitemOpen
  \bibfield  {author} {\bibinfo {author} {\bibfnamefont {Y.}~\bibnamefont
  {Messica}}\ and\ \bibinfo {author} {\bibfnamefont {D.~B.}\ \bibnamefont
  {Gutman}},\ }\bibfield  {title} {\bibinfo {title} {Hall {Coulomb} drag
  induced by electron-electron skew scattering},\ }\href
  {https://doi.org/10.1103/PhysRevB.110.115424} {\bibfield  {journal} {\bibinfo
   {journal} {Phys. Rev. B}\ }\textbf {\bibinfo {volume} {110}},\ \bibinfo
  {pages} {115424} (\bibinfo {year} {2024})}\BibitemShut {NoStop}%
\bibitem [{\citenamefont {Sinitsyn}\ \emph {et~al.}(2007)\citenamefont
  {Sinitsyn}, \citenamefont {MacDonald}, \citenamefont {Jungwirth},
  \citenamefont {Dugaev},\ and\ \citenamefont {Sinova}}]{Sinitsyn2007}%
  \BibitemOpen
  \bibfield  {author} {\bibinfo {author} {\bibfnamefont {N.~A.}\ \bibnamefont
  {Sinitsyn}}, \bibinfo {author} {\bibfnamefont {A.~H.}\ \bibnamefont
  {MacDonald}}, \bibinfo {author} {\bibfnamefont {T.}~\bibnamefont
  {Jungwirth}}, \bibinfo {author} {\bibfnamefont {V.~K.}\ \bibnamefont
  {Dugaev}},\ and\ \bibinfo {author} {\bibfnamefont {J.}~\bibnamefont
  {Sinova}},\ }\bibfield  {title} {\bibinfo {title} {Anomalous {Hall} effect in
  a two-dimensional {Dirac} band: The link between the {Kubo-Streda} formula
  and the semiclassical {Boltzmann} equation approach},\ }\href
  {https://doi.org/10.1103/PhysRevB.75.045315} {\bibfield  {journal} {\bibinfo
  {journal} {Phys. Rev. B}\ }\textbf {\bibinfo {volume} {75}},\ \bibinfo
  {pages} {045315} (\bibinfo {year} {2007})}\BibitemShut {NoStop}%
\bibitem [{\citenamefont {K{\"o}nig}\ and\ \citenamefont
  {Levchenko}(2021)}]{Konig2021}%
  \BibitemOpen
  \bibfield  {author} {\bibinfo {author} {\bibfnamefont {E.~J.}\ \bibnamefont
  {K{\"o}nig}}\ and\ \bibinfo {author} {\bibfnamefont {A.}~\bibnamefont
  {Levchenko}},\ }\bibfield  {title} {\bibinfo {title} {Quantum kinetics of
  anomalous and nonlinear {Hall} effects in topological semimetals},\ }\href
  {https://doi.org/10.1016/j.aop.2021.168492} {\bibfield  {journal} {\bibinfo
  {journal} {Ann. Phys.}\ }\textbf {\bibinfo {volume} {435}},\ \bibinfo {pages}
  {168492} (\bibinfo {year} {2021})}\BibitemShut {NoStop}%
\bibitem [{\citenamefont {Bandurin}\ \emph {et~al.}(2018)\citenamefont
  {Bandurin}, \citenamefont {Shytov}, \citenamefont {Levitov}, \citenamefont
  {Kumar}, \citenamefont {Berdyugin}, \citenamefont {Shalom}, \citenamefont
  {Grigorieva}, \citenamefont {Geim},\ and\ \citenamefont
  {Falkovich}}]{Bandurin2018}%
  \BibitemOpen
  \bibfield  {author} {\bibinfo {author} {\bibfnamefont {D.~A.}\ \bibnamefont
  {Bandurin}}, \bibinfo {author} {\bibfnamefont {A.~V.}\ \bibnamefont
  {Shytov}}, \bibinfo {author} {\bibfnamefont {L.~S.}\ \bibnamefont {Levitov}},
  \bibinfo {author} {\bibfnamefont {R.~K.}\ \bibnamefont {Kumar}}, \bibinfo
  {author} {\bibfnamefont {A.~I.}\ \bibnamefont {Berdyugin}}, \bibinfo {author}
  {\bibfnamefont {M.~B.}\ \bibnamefont {Shalom}}, \bibinfo {author}
  {\bibfnamefont {I.~V.}\ \bibnamefont {Grigorieva}}, \bibinfo {author}
  {\bibfnamefont {A.~K.}\ \bibnamefont {Geim}},\ and\ \bibinfo {author}
  {\bibfnamefont {G.}~\bibnamefont {Falkovich}},\ }\bibfield  {title} {\bibinfo
  {title} {Fluidity onset in graphene},\ }\href
  {https://doi.org/10.1038/s41467-018-07004-4} {\bibfield  {journal} {\bibinfo
  {journal} {Nat. Commun.}\ }\textbf {\bibinfo {volume} {9}},\ \bibinfo {pages}
  {4533} (\bibinfo {year} {2018})}\BibitemShut {NoStop}%
\bibitem [{\citenamefont {Berdyugin}\ \emph {et~al.}(2019)\citenamefont
  {Berdyugin}, \citenamefont {Xu}, \citenamefont {Pellegrino}, \citenamefont
  {Kumar}, \citenamefont {Principi}, \citenamefont {Torre}, \citenamefont
  {Shalom}, \citenamefont {Taniguchi}, \citenamefont {Watanabe}, \citenamefont
  {Grigorieva}, \citenamefont {Polini}, \citenamefont {Geim},\ and\
  \citenamefont {Bandurin}}]{Berdyugin2019}%
  \BibitemOpen
  \bibfield  {author} {\bibinfo {author} {\bibfnamefont {A.~I.}\ \bibnamefont
  {Berdyugin}}, \bibinfo {author} {\bibfnamefont {S.~G.}\ \bibnamefont {Xu}},
  \bibinfo {author} {\bibfnamefont {F.~M.}\ \bibnamefont {Pellegrino}},
  \bibinfo {author} {\bibfnamefont {R.~K.}\ \bibnamefont {Kumar}}, \bibinfo
  {author} {\bibfnamefont {A.}~\bibnamefont {Principi}}, \bibinfo {author}
  {\bibfnamefont {I.}~\bibnamefont {Torre}}, \bibinfo {author} {\bibfnamefont
  {M.~B.}\ \bibnamefont {Shalom}}, \bibinfo {author} {\bibfnamefont
  {T.}~\bibnamefont {Taniguchi}}, \bibinfo {author} {\bibfnamefont
  {K.}~\bibnamefont {Watanabe}}, \bibinfo {author} {\bibfnamefont {I.~V.}\
  \bibnamefont {Grigorieva}}, \bibinfo {author} {\bibfnamefont
  {M.}~\bibnamefont {Polini}}, \bibinfo {author} {\bibfnamefont {A.~K.}\
  \bibnamefont {Geim}},\ and\ \bibinfo {author} {\bibfnamefont {D.~A.}\
  \bibnamefont {Bandurin}},\ }\bibfield  {title} {\bibinfo {title} {Measuring
  {Hall} viscosity of graphene's electron fluid},\ }\href
  {https://doi.org/10.1126/science.aau0685} {\bibfield  {journal} {\bibinfo
  {journal} {Science}\ }\textbf {\bibinfo {volume} {364}},\ \bibinfo {pages}
  {162} (\bibinfo {year} {2019})}\BibitemShut {NoStop}%
\bibitem [{\citenamefont {Talanov}\ \emph {et~al.}()\citenamefont {Talanov},
  \citenamefont {Waissman}, \citenamefont {Hui}, \citenamefont {Skinner},
  \citenamefont {Watanabe}, \citenamefont {Taniguchi},\ and\ \citenamefont
  {Kim}}]{Waissman2024}%
  \BibitemOpen
  \bibfield  {author} {\bibinfo {author} {\bibfnamefont {A.}~\bibnamefont
  {Talanov}}, \bibinfo {author} {\bibfnamefont {J.}~\bibnamefont {Waissman}},
  \bibinfo {author} {\bibfnamefont {A.}~\bibnamefont {Hui}}, \bibinfo {author}
  {\bibfnamefont {B.}~\bibnamefont {Skinner}}, \bibinfo {author} {\bibfnamefont
  {K.}~\bibnamefont {Watanabe}}, \bibinfo {author} {\bibfnamefont
  {T.}~\bibnamefont {Taniguchi}},\ and\ \bibinfo {author} {\bibfnamefont
  {P.}~\bibnamefont {Kim}},\ }\href@noop {} {\bibinfo {title} {Observation of
  electronic viscous dissipation in graphene magneto-thermal transport}},\
  \Eprint {https://arxiv.org/abs/2406.13799} {arXiv:2406.13799} \BibitemShut
  {NoStop}%
\bibitem [{\citenamefont {Zeng}\ \emph {et~al.}()\citenamefont {Zeng},
  \citenamefont {Guo}, \citenamefont {Ghosh}, \citenamefont {Watanabe},
  \citenamefont {Taniguchi}, \citenamefont {Levitov},\ and\ \citenamefont
  {Dean}}]{Dean2024}%
  \BibitemOpen
  \bibfield  {author} {\bibinfo {author} {\bibfnamefont {Y.}~\bibnamefont
  {Zeng}}, \bibinfo {author} {\bibfnamefont {H.}~\bibnamefont {Guo}}, \bibinfo
  {author} {\bibfnamefont {O.~M.}\ \bibnamefont {Ghosh}}, \bibinfo {author}
  {\bibfnamefont {K.}~\bibnamefont {Watanabe}}, \bibinfo {author}
  {\bibfnamefont {T.}~\bibnamefont {Taniguchi}}, \bibinfo {author}
  {\bibfnamefont {L.~S.}\ \bibnamefont {Levitov}},\ and\ \bibinfo {author}
  {\bibfnamefont {C.~R.}\ \bibnamefont {Dean}},\ }\href@noop {} {\bibinfo
  {title} {Quantitative measurement of viscosity in two-dimensional electron
  fluids}},\ \Eprint {https://arxiv.org/abs/2407.05026} {arXiv:2407.05026}
  \BibitemShut {NoStop}%
\bibitem [{\citenamefont {Messica}\ \emph {et~al.}(2025)\citenamefont
  {Messica}, \citenamefont {Levchenko},\ and\ \citenamefont
  {Gutman}}]{StokesMathematicaData}%
  \BibitemOpen
  \bibfield  {author} {\bibinfo {author} {\bibfnamefont {Y.}~\bibnamefont
  {Messica}}, \bibinfo {author} {\bibfnamefont {A.}~\bibnamefont {Levchenko}},\
  and\ \bibinfo {author} {\bibfnamefont {D.~B.}\ \bibnamefont {Gutman}},\
  }\href {https://doi.org/10.5281/zenodo.15682590} {\bibinfo {title}
  {Mathematica code for calculation of pressure field and streamlines}},\
  \bibinfo {howpublished} {Zenodo, https://doi.org/10.5281/zenodo.15682590}
  (\bibinfo {year} {2025})\BibitemShut {NoStop}%
\end{thebibliography}%

\end{document}